\documentclass[useAMS,usenatbib]{mn2e}
\usepackage{threeparttable}
\usepackage{epsfig}
\usepackage[linktocpage=true]{hyperref}
\hypersetup{colorlinks=true,citecolor=blue,linkcolor=blue,filecolor=black,runcolor=black}
\usepackage{breakurl} 
\usepackage{times}
\usepackage{amsmath}

\newcommand{\kmpers}{\mbox{km~s$^{-1}$}}
\newcommand{\Kkmpers}{\mbox{K~	km~s$^{-1}$}}
\newcommand{\xcounits}{\mbox{cm$^{-2}$ (K km s$^{-1}$)$^{-1}$}}

\newcommand{\xco}{\mbox{$X_{\rm CO}$}}

\newcommand {\aj}{AJ}
\newcommand {\aap}{A\&A}
\newcommand {\aaps}{A\&AS}
\newcommand {\apj}{ApJ}
\newcommand {\apjs}{ApJS}
\newcommand {\apjl}{ApJL}
\newcommand {\mnras}{MNRAS}
\newcommand {\pasp}{PASP}
\newcommand {\pasj}{PASJ}

\newcommand {\araa}{ARA\&A}
\newcommand {\qjras}{QJRAS}

\begin{document}

\title[$I_{\rm CO}$-$A_V$ Relationship in the Magellanic Clouds]{The Relationship between CO Emission and Visual Extinction Traced by Dust Emission in the Magellanic Clouds}

\author[Cheoljong Lee et al.]{Cheoljong Lee,$^{1}$\thanks{cl5ju@virginia.edu}
 Adam K. Leroy,$^{2,3}$ Scott Schnee,$^{3}$ Tony Wong,$^{4}$ Alberto D. Bolatto,$^{5}$ 
\newauthor Remy Indebetouw,$^{1,3}$ Monica Rubio$^{6}$\\
$^{1}$Department of Astronomy, University of Virginia, Charlottesville, VA 22904, USA\\
$^{2}$Department of Astronomy, The Ohio State University, 140 West 18th Avenue, Columbus, OH 43210\\
$^{3}$National Radio Astronomy Observatory, Charlottesville, VA 22903, USA\\
$^{4}$Astronomy Department, University of Illinois, Urbana, IL 61801, USA\\
$^{5}$Department of Astronomy, University of Maryland, College Park, MD 20742, USA\\
$^{6}$Departamento de Astronom\'{i}a, Universidad de Chile, Casilla 36-D, Santiago, Chile}

\date{Accepted 0000 December 00. Received 0000 December 00; in original form 0000 October 00}
\pagerange{\pageref{firstpage}--\pageref{lastpage}} \pubyear{0000}
\maketitle

\begin{abstract}
To test the theoretical understanding that finding bright CO emission depends primarily on dust shielding,
we investigate the relationship between CO emission ($I_{\rm CO}$) and the amount of dust 
(estimated from IR emission and expressed as ``$A_V$'') across the Large Magellanic Cloud, the Small Magellanic Cloud, 
and the Milky Way. 
We show that at our common resolution of 10 pc scales, $I_{\rm CO}$ given a fixed line-of-sight $A_V$ is similar across all three systems despite the difference in metallicity.
We find some evidence for a secondary dependence
of $I_{\rm CO}$ on radiation field; in the LMC, $I_{\rm CO}$ at a given
$A_V$ is smaller in regions of high $T_{\rm dust}$, perhaps because of an increased photodissociating radiation field.
We suggest a simple but useful picture in which the CO-to-H$_2$ conversion factor (\xco) depends on two 
separable factors: (1) the distribution of gas column densities, which maps to an extinction 
distribution via a dust-to-gas ratio; and (2) the dependence of $I_{\rm CO}$ on $A_V$. Assuming that the probability distribution
function (PDF) of local Milky Way clouds is universal, this approach predicts a dependence of $\xco$ on $Z$ between $Z^{-1}$ and $Z^{-2}$
above about a third solar metallicity. Below this metallicity, CO emerges from only the high column density parts of the cloud
and so depends very sensitively on the adopted PDF and the H$_2$/{\sc Hi} prescription. The PDF of low metallicity 
clouds is thus of considerable interest and the uncertainty associated with even an ideal prescription for \xco\ at very low metallicity
will be large.
\end{abstract}


\section{Introduction}
\label{sec:intro}

As the immediate reservoir for star formation, the molecular interstellar medium (ISM) plays a key role in the evolution of galaxies. 
Unfortunately, the majority of cold molecular hydrogen (H$_2$) in typical clouds is invisible in emission due to the fact that the H$_2$ molecule has low mass 
and therefore requires high temperatures (the lowest level corresponds to $E/k \approx 510$~K) to excite its rotational transitions \citep{KENNICUTT12}.
As a result, astronomers employ a suite of more observationally accessible tracers of H$_2$ to study molecular clouds. 
Low $J$ rotational transitions of CO represent the most accessible and commonly used such tracers, especially in external galaxies. 
The abundance, pervasiveness, and brightness of CO make it a useful tracer, but it is not perfectly coextant with H$_2$ and
the relationship between CO emission and H$_2$ column density (so called ``CO-to-H$_2$ conversion factor''; \xco) is both expected
and observed to vary systematically as local conditions change \citep[see the review by][]{BOLATTO13}. 
Consequently, it is important to understand the physical origins of CO emission and their implications for the use of CO emission to trace H$_2$.

Coarsely, variations in $\xco$ are twofold. The regions in a molecular cloud where CO and H$_2$ exist are not perfectly
matched, with a layer of CO-poor H$_2$ extending beyond the region at which the dominant form of carbon changes from CO to C{\sc ii}
\citep[{\rm e.g.,}][]{MB88, VB88}. Theoretically, the amount of dust shielding ($A_V$) between the CO-C{\sc ii} transition layer and the H$_2$-H{\sc i} transition layer
is estimated to be almost constant \citep{WOLFIRE10}. This is also seen in PDR calculations \citep{BELL06} and numerical simulations \citep{GLOVER11,SHETTY11},
where $I_{\rm CO}$ exhibits a clear dependence on $A_V$.
In a low metallicity cloud the dust-to-gas ratio is also low, so that achieving some fixed $A_V$ requires 
a much larger column of gas than at high metallicity. Therefore this intermediate region of H$_2$ without
much associated CO becomes very large in terms of total gas content. As a result, one expects to find less 
CO emission per unit H$_2$ in clouds with a low metallicity. This gives rise to a dust-to-gas ratio, and thus metallicity, dependent term in $\xco$. 
Meanwhile, within region where CO is abundant, the line is usually optically thick. This leads the ratio of CO emission
to gas mass in this region to depend on density, temperature, and potentially other dynamical factors \citep[e.g., see][]{MB88,DOWNES98,NARAYANAN12}.

In this paper, we focus on the first part of the problem, the metallicity-dependent term in $\xco$. We focus on the relationship
between dust abundance along the line of sight and the brightness of CO emission (the ``$I_{\rm CO}$-$A_V$'' relation) as a way to 
explore the physics of CO emission and its utility as a tracer of H$_2$. Because theoretical models highlight the key role of dust shielding in setting the extent
of bright CO emission, our hypothesis is that across diverse environments we will often find about the same amount of CO emission
per unit dust shielding.

Highly resolved (sub-pc) observations of individual nearby clouds have explored the $I_{\rm CO}$-$A_V$ relationship in detail \citep{LOMBARDI06, PINEDA08, PINEDA10}.
These provide strong observational support for the crucial role of $A_V$ in determining the amount CO emission. These high resolution
studies also reveal distinct regimes in the relationship, such as an extinction threshold below which CO emission is faint or absent, a linear rise of $I_{\rm CO}$
at intermediate extinctions, and evidence for saturation at high extinctions. The relationship has not been explored as much outside the Milky Way
because of the coarse physical resolution in most CO and dust maps of other galaxies. However, studying a molecular complex in the Small Magellanic Cloud (SMC) at
$\approx 10$~pc resolution, \cite{LEROY09} did observe an $I_{\rm CO}$-$A_V$ relationship that resembled that for a Milky Way cloud (their Figure~7).

Several new data sets make it possible to revisit the relationship between dust and CO emission in the Magellanic clouds over a much wider area.
The Magellanic Mopra Assessment (MAGMA) \citep{WONG11} obtained high spatial resolution ($\sim 10$~pc) CO $J = 1 \to 0$ data 
across most areas of bright CO emission in the LMC. A new APEX survey of the southwest part of the SMC (Rubio et al., in prep.) provides similar 
coverage in that galaxy; we also use previously published CO $J = 2 \to 1$ and $J = 1 \to 0$ maps of the N83 complex in the SMC Wing \citep{BOLATTO03}.
Key projects by {\em Spitzer} \citep[Surveying the Agents of a Galaxy's Evolution, SAGE][]{MEIXNER06} and {\em Herschel} \citep[HERschel Inventory of The 
Agents of Galaxy Evolution, HERITAGE][]{MEIXNER10, MEIXNER13} allow us to model infrared (IR) emission to estimate the line of sight extinction.
This provides a handle on the total extinction (or dust column) through a part of the galaxy, which offers an imperfect but observationally accessible 
analog to the physically crucial shielding of material from the interstellar radiation field (ISRF). Clearly, the relation between the total dust column
(expressed by ``$A_V$'') and the degree of shielding towards an average CO molecule depends on geometry, but even in simulations
$A_V$ and real shielding appear closely related \citep[{\rm e.g., see}][]{GLOVER11}.  Thus, we now have a handle on CO emission and dust
column at $\approx 10$~pc resolution over a matched area for the two nearest \citep[$\sim$ 50~kpc for the LMC and 60~kpc for the SMC][]{KELLER06} 
star-forming low-metallicity galaxies \citep[$\sim$ 1/2~$Z_\odot$ for the LMC and 1/5~$Z_\odot$ for the SMC][]{WESTERLUND97}.

Combining these data on the Magellanic Clouds with Milky Way data from {\em Planck} and the \citet{DAME01} CO survey, we are able to ask how the 
CO intensity, $I_{\rm CO}$, at a given line of sight extinction, $A_V$,  compares between the Milky Way, the LMC, and SMC. Do successive steps of a 
factor of $\approx 2$ in metallicity have a visible impact on $I_{\rm CO}$ at a given $A_V$ or is the amount of dust shielding  alone the key parameter?
We also search for secondary factors affecting $I_{\rm CO}$ at fixed $A_V$, with the most obvious candidate being the interstellar radiation field, which is 
directly traced by the dust temperature. This might be expected to influence the amount of CO emission at a given line of sight extinction in two ways:
first lowering the amount of CO emission by raising the number of dissociating photons and so requiring more dust shielding for the transition from
{\sc Cii} to CO. Second, perhaps increasing the temperature of the CO and so increasing $I_{\rm CO}$.

In the second part of the paper, we explore the implications of a universal $I_{\rm CO}$-$A_V$ relation for the CO-to-H$_2$ conversion 
factor, $\xco$. If $I_{\rm CO}$ is largely set once $A_V$ is known, then the distribution of $A_V$ becomes the key factor to predict CO emission.
This, in turn, depends on the probability distribution function (PDF) of gas column densities and the gas-to-dust ratio. The PDF of individual clouds in the 
Milky Way has been the study of significant quantitative study in recent years \citep[see][and following]{KAIN09}. We combine these results with
our estimates of the $I_{\rm CO}$-$A_V$ relation to make an empirically-driven estimate of how \xco\ depends on metallicity.

\section{Data and Modelling}
\label{sec:data}

We aim to compare CO emission to the line-of-sight extinction, estimated from IR emission, on the scale of individual clouds ($\sim$ 10~pc). To do
so, we assemble matched-resolution CO (Section \ref{sec:co}) and infrared (IR) emission maps (Section \ref{sec:ir}) for the LMC, SMC, and Milky Way. 
We use the IR data to estimate the line of sight extinction (Section \ref{sec:av}) and so estimate $I_{\rm CO}$ as a function of $A_V$.

\subsection{CO Data}
\label{sec:co}

\subsubsection{LMC}

We use the second release of the Magellanic MOPRA Assessment \citep[MAGMA][]{WONG11} survey\footnote{\url{http://mmwave.astro.illinois.edu/magma/DR2b/}}
to generate an integrated intensity (``moment 0'') map of CO emission from the LMC. MAGMA used the 22-m MOPRA telescope to observe the CO $J
= 1 \to 0$ transition toward molecular clouds identified from NANTEN surveys \citep{FUKUI99, FUKUI08}. MAGMA has an
angular resolution of 45\arcsec, $\approx 10$~pc at the distance of the LMC. The mean rms brightness temperature of the MAGMA cube in a single channel (0.5~\kmpers) is 0.3~K \citep{WONG11}. 
We generate the integrated intensity map by directly integrating the cube along the whole velocity axis ($180~\kmpers \le V_{\rm LSR} \le 320~\kmpers$).

For both theoretical and practical reasons, our analysis will treat $A_V$ as the independent variable.
$A_V$ is expected to set the amount of CO emission and IR emission is detected at higher signal to noise than CO
throughout the Magellanic Clouds. Reflecting this, we work with integrated intensity derived from a broad
velocity window that will include any CO emission from the LMC. This directly integrated intensity map is not clipped
or masked and so includes both positive and negative values. The advantage of this approach is that by averaging together 
many spectra at the same $A_V$ we can recover a mean $I_{\rm CO}$ that is too faint to be detected in an individual line of sight.
A corollary of this approach is that we are sensitive to small zero point offsets in the data. Therefore, 
we subtract a constant baseline from the MAGMA data cube pixel-by-pixel, with the value determined from the median 
intensity of each line of sight from the signal-free edge channels at the edge of the data cube. The median offset in the zero level 
is very small, corresponding to only $\sim$ 1.2~mK and only important because some of our analysis focuses on faint regions. 
In Appendix~\ref{sec:appendix_baseline} we show the quantitative effects of varying our baseline treatment, 
which is very minor compared to other uncertainties in the analysis.

\subsubsection{SMC}

For the SMC, we use a new APEX survey of CO $J = 2 \to 1$ emission from the southwest region of the SMC data (P.I.: Rubio, Rubio et al., in prep.).
The angular resolution of the data is $\sim$ 28\arcsec, corresponding to $\approx$ 8~pc at the distance of the SMC,
and rms noise is 0.3~K in each 0.1~\kmpers channel. These data target the southwest part of the SMC bar, which contains most of the ongoing
star formation and molecular gas in the SMC. We also use SEST observation of CO $J = 2 \to 1$ and  $J = 1 \to 0$ emission from 
the star forming complex N83 located in the wing of the SMC \citep{BOLATTO03}. These data have angular resolution of 38\arcsec ($J = 2 \to 1$) 
and 55\arcsec ($J = 1 \to 0$) and rms noise 0.1~K in a 0.25~\kmpers channel for both lines.

As in the LMC, we directly integrate these data along the velocity axis to generate integrated CO intensity maps of the southwest SMC,
picking a velocity range that covers the whole region and carrying out no other ``masking.'' The velocity ranges of integration are 
$80~\kmpers \le V_{\rm LSR} \le 160~\kmpers$ for the southwest SMC and  $140~\kmpers \le V_{\rm LSR} \le 187~\kmpers$ for N83.

\subsection{Infrared Maps}
\label{sec:ir}

We use observations of IR dust emission in the LMC and the SMC at four different wavelengths from the HERITAGE survey: 
100 and 160~$\mu$m images from the PACS instrument, and 250 and 350~$\mu$m images from the SPIRE instrument.
The  angular resolutions of these maps are 7.7\arcsec (100~$\mu$m), 12\arcsec (160~$\mu$m), 18\arcsec (250~$\mu$m), and 25\arcsec (350~$\mu$m),
so that the CO data set the limiting resolution for our analysis. 

Before proceeding to our analysis, we convolved the IR maps in the LMC to the 45\arcsec
resolution of the MAGMA CO map. This is done by first using the convolution kernels of \citet{ANIANO11} to convolve HERITAGE maps to
a common PSF (we used the {\em Spitzer} 160$\mu$m for comparison with {\em Spitzer} work) and then degrade them together to
the $45\arcsec$ resolution of the MAGMA CO data. We place all LMC data on the same astrometric grid, which has  pixel spacing of 15.6\arcsec.

We take a similar approach to match the resolutions and grids of IR maps in the SMC. Here we match the {\em Herschel} data to the 
28\arcsec\ resolution of the CO $J = 2 \to 1$ transition in the southwest region of the SMC, to the 38\arcsec\ resolution for the CO $J = 2 \to 1$ transition in N83, 
and to the 55\arcsec\ for the $J = 1 \to 0$ transition in N83. Again all data are placed on a shared astrometric grid.

The 1-$\sigma$ noises of the LMC IR maps at our working resolution (45\arcsec) are $\sigma_{100} \sim $ 2.3~MJy~sr$^{-1}$,  $\sigma_{160} \sim $ 2.4~MJy~sr$^{-1}$,
$\sigma_{250} \sim $ 0.88~MJy~sr$^{-1}$, and $\sigma_{350} \sim $ 0.48~MJy~sr$^{-1}$. In our analysis we consider only regions with
intensity at least three times these values in each band. We refer to the ``LMC field'' as the region that satisfies this S/N ratio cut in the IR maps, 
while the ``MAGMA field'' is defined as the region in the LMC field where the MAGMA survey mapped in CO. The MAGMA field is a subset of the LMC field.

The noises in the SMC IR maps are similar but vary with resolution. We estimate the noise at each working resolution and again 
mask regions below 3-$\sigma$ in the IR maps. As above, the ``SMC field'' is the region where IR maps have values greater than 3-$\sigma$ at 
high resolution, while the ``APEX field'' and ``SEST field'' are the regions within the SMC field where APEX telescope and SEST telescope 
mapped CO emission.

Note that a first order (linear) baseline has been already subtracted from the IR maps in the HERITAGE release to remove emission not associated with the Magellanic
Clouds themselves from the maps \citep{MEIXNER13}. We do not apply any further correction to account for Milky Way foreground dust emission, though we do model a 
level of uncertainty in this subtraction by including a zero point uncertainty in our Monte-Carlo analysis (Section~\ref{sec:unc}).

\subsection{Estimation of the ``$A_{V}$" Map}
\label{sec:av}

We use the \emph{Herschel} IR emission maps to estimate ``$A_{V}$'' along each line of sight through the LMC and the SMC. Here $A_{V}$ refers to visual
extinction, measured in magnitudes estimated from the optical depth at 160~$\mu$m, $\tau_{160}$. We calculate this by fitting a modified blackbody to the 
measured IR intensities (Section \ref{sec:modbb}) and then converting the dust optical depth to visual extinction following an empirical scaling derived from 
the Milky Way (Section \ref{sec:tau_to_av}). Because the dust emission is optically thin, ``$A_{V}$" measured in this way will probe material along the whole 
line of sight and averaged over the substantial beam of the {\em Herschel} data. This is similar to extinction mapping for nearby molecular clouds, which uses sources
behind the cloud to create a large scale map. It differs from true extinction mapping using internal sources in the Magellanic Clouds ({\rm e.g.}, stars), which 
will measure the extinction only part of the way through the galaxy.

\subsubsection{Modified Blackbody Fit}
\label{sec:modbb}

We assume that the dust along a line of sight can be described as an optically thin ($\tau \ll 1$) greybody at an equilibrium temperature $T_{dust}$,
and wavelength dependence of dust optical depth is a power law with spectral index $\beta$, {\rm i.e.}, $\tau_\lambda \propto \lambda^{-\beta}$
\citep[{\rm e.g.,}][]{DL84}. In this case the optical depth at 160~$\mu$m ($\tau_{160}$) is given by

\begin{equation}
\label{tau160}
\tau_{160} = \frac{I_{160}}{B_{\nu}(T_{dust}, 160~\micron)},
\end{equation}

\noindent where $I_{160}$ is the observed 160~$\mu$m intensity and $B_{\nu}$($T_{dust}$, $\lambda$) is the intensity of a blackbody of temperature $T_{dust}$ at wavelength $\lambda$. 

Because we have dust emission intensities measured at four different wavelengths, in principle we can fit for the three unknowns $\beta$, $\tau_{160}$ and  $T_{dust}$.
Instead we adopt a fixed $\beta = 1.5$ as our fiducial value and fit for two unknowns $\tau_{160}$ and  $T_{dust}$ by minimizing the $\chi^2$ from observed IR intensities and the 
model IR intensities, taking the colour correction\footnote{See \url{http://herschel.esac.esa.int/twiki/pub/Public/PacsCalibrationWeb/cc_report_v1.pdf} for PACS colour correction and 
\url{http://herschel.esac.esa.int/hcss-doc-11.0/load/spire_drg/html/ch05s07.html} for SPIRE colour correction.} for each filter into account. We fix $\beta$ in order to minimize the 
uncertainties on the $T_{dust}$ (and thus $\tau_{160}$) arising from the fact that $\beta$ and $T_{dust}$ are somewhat degenerate in $\chi^2$ space \citep{DUPAC03}.  The adopted $\beta$ above is a reasonable
description of the integrated SED of the LMC \citep{BERNARD08, GORDON10, PLANCK11} and the SMC \citep{STANIMIROVIC00,AGUIRRE03,LEROY07}, and is
intermediate in the range of plausible astrophysical values, $1.0 < \beta < 2.0$ (\citealt{DL84}, \citealt{SFD98}).

\subsubsection{Conversion to $A_{V}$}
\label{sec:tau_to_av}

Our modified blackbody fit yields the optical depth at 160~$\mu$m, $\tau_{160}$. We wish to phrase our analysis in terms of the line of
sight $V$-band extinction, $A_{V}$, which is the conventional unit expressing shielding in discussion of PDR regions. 
We translate $\tau_{160}$ to $A_{V}$ via

\begin{equation}
\label{av_tau160}
A_{V} \sim 2200~\tau_{160}~.
\end{equation}

\noindent We arrive at this conversion in several ways and take the differences among the estimates to indicate the uncertainty in the conversion.
First, following \citet{LEROY09}, we infer $\tau_{160} = 2.44 \times10^{-25}$~$N$(H{\sc i}) from DIRBE/FIRAS observations of the Galactic
diffuse H{\sc i} \citep{1996A&A...312..256B}. Then, we adopt the relation between colour excess of Solar Neighborhood stars and H{\sc i} column density
from Ly $\alpha$ absorption, $E(B - V) = N(\rm H)/5.8 \times$ 10$^{21}$~cm$^{-2}$ \citep{BOHLIN78}. Taking a Galactic $R_V = 3.1$, we estimate $A_{V} \sim 2190~\tau_{160}$. 

We compare this number to the \citet{SFD98} conversion used to estimate Galactic extinction maps from IRAS data. After correction to
a fiducial dust temperature, they find $E(B-V) = \left( 0.016 \pm 0.004 \right)~I_{100}^T$, where $I_{100}^T$ is the 100~$\mu$m intensity
after correction to a fixed 18.2~K temperature, and  $E(B-V)$ is reddening measurement of  background elliptical galaxies.
Again taking $R_V = 3.1$ and assuming $\beta = 2$ for Galactic dust\footnote{In comparing our $\tau_{160}$-$A_V$ conversion to other works,
we use $\beta$ assumed in each study rather than our fiducial $\beta = 1.5$.}, the \citet{SFD98} conversion corresponds to $A_{V} \sim
1939~\tau_{160}$. More recently, \citet{PLANCK13_DUST} find systematically higher dust temperature and thus lower dust optical depth at high Galactic latitude
using IRAS and {\em Planck} data. This is mainly due to the observed dust emission SED being flatter ($\beta \approx 1.59$) than previous studies.
This leads to a higher $E(B-V)$ per dust optical depth ($\tau$), where the former is from SDSS quasars. Adopting $R_V = 3.1$ and $\beta$ from their study, 
the conversion is $A_{V} \sim 3246~\tau_{160}$.

This conversion can also be related directly to the mass absorption coefficient of dust, $\kappa$, following \citet{HILDEBRAND83}. We
consider $\tau_{160} = \Sigma_{dust} \kappa_{160}$. In the Milky Way, \citet{HILDEBRAND83} suggests $\kappa_{250} \approx
10$~cm$^{2}$~g$^{-1}$, which we convert to 160~$\mu$m assuming $\beta=2$. Then taking the relation between $E(B-V)$ and $N(\rm H)$ above
and assuming a dust-to-gas mass ratio of $\sim 1$-to-$150$ \citep{DRAINE07} and $R_{V} = 3.1$, we arrive at $A_{V} \sim 1443~\tau_{160}$.

To synthesize, calculations based on Milky Way studies suggest $A_{V} = (1400 - 3200)~\tau_{160}$, with the mean conversion corresponding to $A_{V}
\approx 2200~\tau_{160}$, which is the conversion we use in this paper. The fractional uncertainty above approximately matches that of \citet{SFD98}, 
and we fold it in to our Monte-Carlo analysis to account for the uncertainties in the conversion from $\tau_{160}$ to $A_V$ (See Section \ref{sec:unc}). 
 
\begin{figure}
\includegraphics[width=0.99\columnwidth]{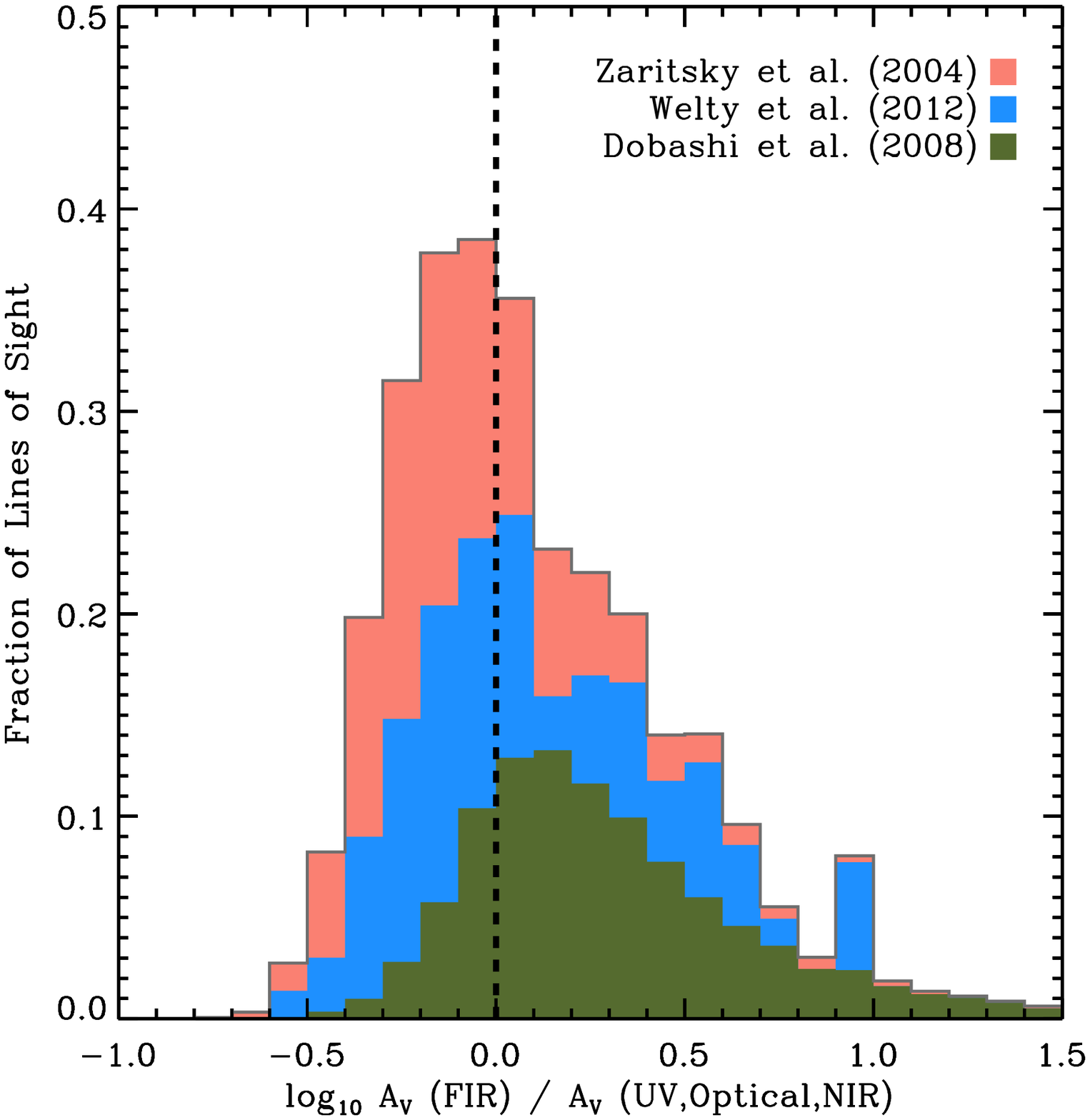}
\caption{\label{avtoir} Comparison of our FIR-based $A_V$ estimate to $A_V$ inferred from stellar photometry \citep{ZARITSKY04}, UV spectroscopy \citep{WELTY12}, 
and NIR colour excess \citep{DOBASHI08} in the LMC. The log ratio of the former to the latter ($\log_{10} {A_{V}^{\rm FIR}/A_{V}^{\rm UV,Opt,NIR}}$) is shown on the $x$-axis,
and the fraction of lines of sight is shown on the $y$-axis. The histogram for \citet{ZARITSKY04} is shown on top of the \citet{WELTY12} histogram,
which is on top of the \citet{DOBASHI08} histogram. We expect the \citet{ZARITSKY04} and \citet{WELTY12} $A_V$ estimates to sample only about half of the
LMC, on average, and so scale these estimates by a factor of two in the plot. The dotted line indicates a ratio of unity. This lies close to the center of the sum of three histograms,
indicating an overall agreement between our approach to estimate $A_V$ from the FIR and direct measurements, though there is substantial scatter and systematic effects remain.
(A colour version of this figure is available in the online journal.)}
\end{figure}
 
There are several direct measurements of $E(B-V)$ and $A_{V}$ in the LMC. Our $\tau_{160}$ map has complete coverage and high signal-to-noise compared to 
these maps, so they do not offer a viable replacement, but we use them to check our adopted conversion from $\tau_{160}$ to $A_V$ (Equation \ref{av_tau160}).
We compare to three data sets: the $A_{V}$ map estimated by \citet{ZARITSKY04}, which is based on photometry of individual ``hot'' stars; the compilation of 
spectroscopic measurements from \citet{WELTY12}; and the $A_{V}$ map inferred from the near-IR (NIR) colour excess method by \citet{DOBASHI08} using 
2MASS data. In all cases, we restrict the comparison to regions that have $A_{V} > 0.5$~mag in our FIR-based $A_{V}$ map.
\citet{ZARITSKY04} and \citet{WELTY12} report extinction to LMC sources that we expect, on average, to lie halfway through the galaxy. To compare to our
map, which samples the whole column, we multiply those data by a factor of two to account for the difference in geometry. The \citet{DOBASHI08} map already
accounts for the distribution of stars along the line of sight and attempts to report an integrated extinction. Figure \ref{avtoir} shows the resulting comparison as a
histogram of the ratio between our map and these other estimates.

Our FIR-based map yields lower $A_V$, on average than stellar photometry or UV spectroscopy, after the factor of $2$ scaling of the Optical and UV estimates.
We find median $A_{V}^{\rm FIR}/A_{V}^{\rm Opt,NIR} \approx 0.75$ for the \citet{ZARITSKY04} data and median $A_{V}^{\rm FIR}/A_{V}^{\rm UV} \approx 0.95$ 
comparing to the \citet{WELTY12} data. In both cases we observe significant scatter, $\approx 0.24$~dex in the \citet{ZARITSKY04} case and $\approx 0.33$~dex comparing to
\citet{WELTY12}. Much of the large scatter likely reflects the internal geometry of the LMC or, for \citet{WELTY12}, the difference between our large
beam and the single pencil beam sampled by their spectroscopy.  On the other hand, the FIR-based $A_V$ yields somewhat higher $A_V$ than the NIR color excess method.
We find median $A_{V}^{\rm FIR}/A_{V}^{\rm NIR} \approx 1.6$, again with large scatter (0.33~dex). Altogether, the sum of three histograms is approximately centered at a 
ratio of unity (dotted line). Given the stark difference in approaches to estimate the extinction we view these comparisons as reasonable confirmation of our adopted 
$A_{V}/\tau_{160}$ and overall approach. We take them to confirm the $\approx 50\%$ systematic uncertainty in $A_{V}/\tau_{160}$ discussed above.
 
\subsubsection{Estimating the Uncertainties in $\tau_{160}$ and $A_{V}$}
\label{sec:unc}

We are primarily interested in the average relation between $A_{V}$ and $I_{\rm CO}$ in the Magellanic Clouds. To estimate the uncertainties involved in the $A_V$ 
portion of this relation, we adopt a Monte-Carlo approach. We use the LMC data as a basis to repeatedly re-calculate our $A_{V}$ map while varying our assumptions 
across their plausible range. Each time that we do so, we add a new realization of the statistical noise to the data. We begin with the true LMC maps. We then
add Gaussian noise with magnitude matched to the measured statistical noise to each map. For each new map, we also vary the zero point of the maps within 
the estimated uncertainty. Doing so, we generate 100 new maps that could be realistic observations if the observed LMC maps were indeed the true intensity on the sky. 
For each set of these noise-added maps, we generated 10 sets of $T_{dust}$ and $\tau_{160}$ maps following the $\chi^2$ minimization described in Section \ref{sec:modbb}.
For these maps instead of fixing $\beta$ at $1.5$, we fixed it randomly at a value within its plausible ranges, $1 < \beta < 2$. In the end, we have 1000
maps with a spread that captures our true uncertainty regarding the derivation of $A_{V}$, except the conversion from $\tau_{160}$ to $A_V$.
We take this into account by randomly taking a plausible conversion from  $\tau_{160}$ to $A_V$ as determined in Section \ref{sec:tau_to_av}, 
$A_{V}/\tau_{160} = (1400 - 3200)$, resulting in 1000 $A_{V}$ maps. 

Based on these calculations, we calculate the a 1-$\sigma$ fractional scaling uncertainty associated with $T_{dust}$, 10\%, and with $A_V$, 45\%. That is, the whole
map is uncertain by this amount due to zero-point uncertainties and methodological decisions. This is a correlated error that will adjust the entire map. 
The exercise and the scatter would be almost the same for the SMC, so we take these errors as representative of both galaxies.
In the following figures discussing $I_{\rm CO}$-$A_V$ relationships, we show this error estimate for $A_V$ as the horizontal error bars in the bottom-right corners to represent the typical uncertainty in $A_V$. 

\begin{figure*}
\includegraphics[width=0.49\linewidth]{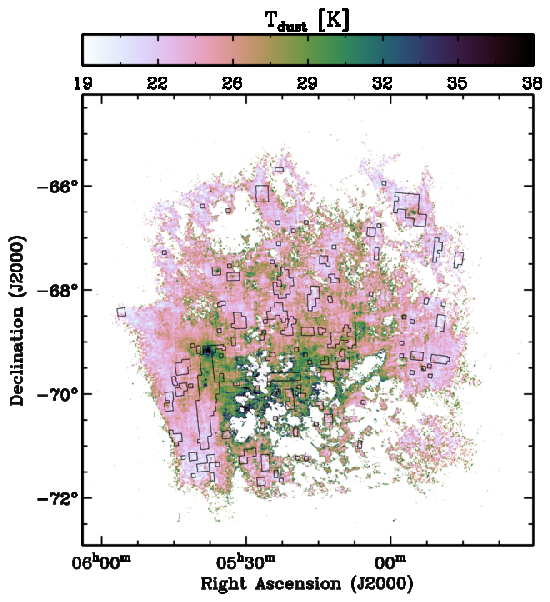} \includegraphics[width=0.49\linewidth]{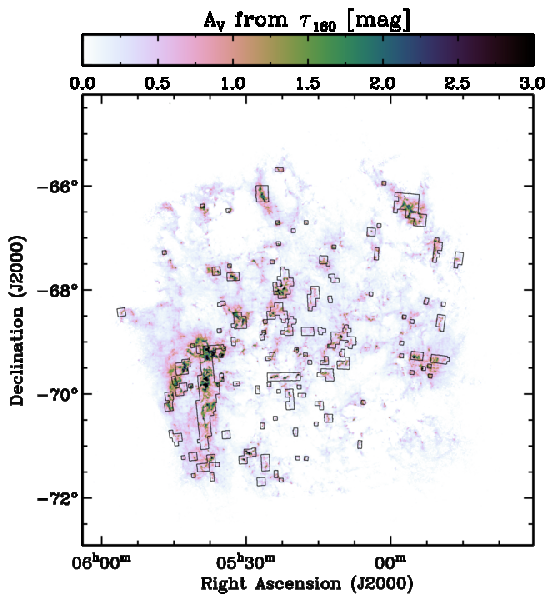} 
\includegraphics[width=0.49\linewidth]{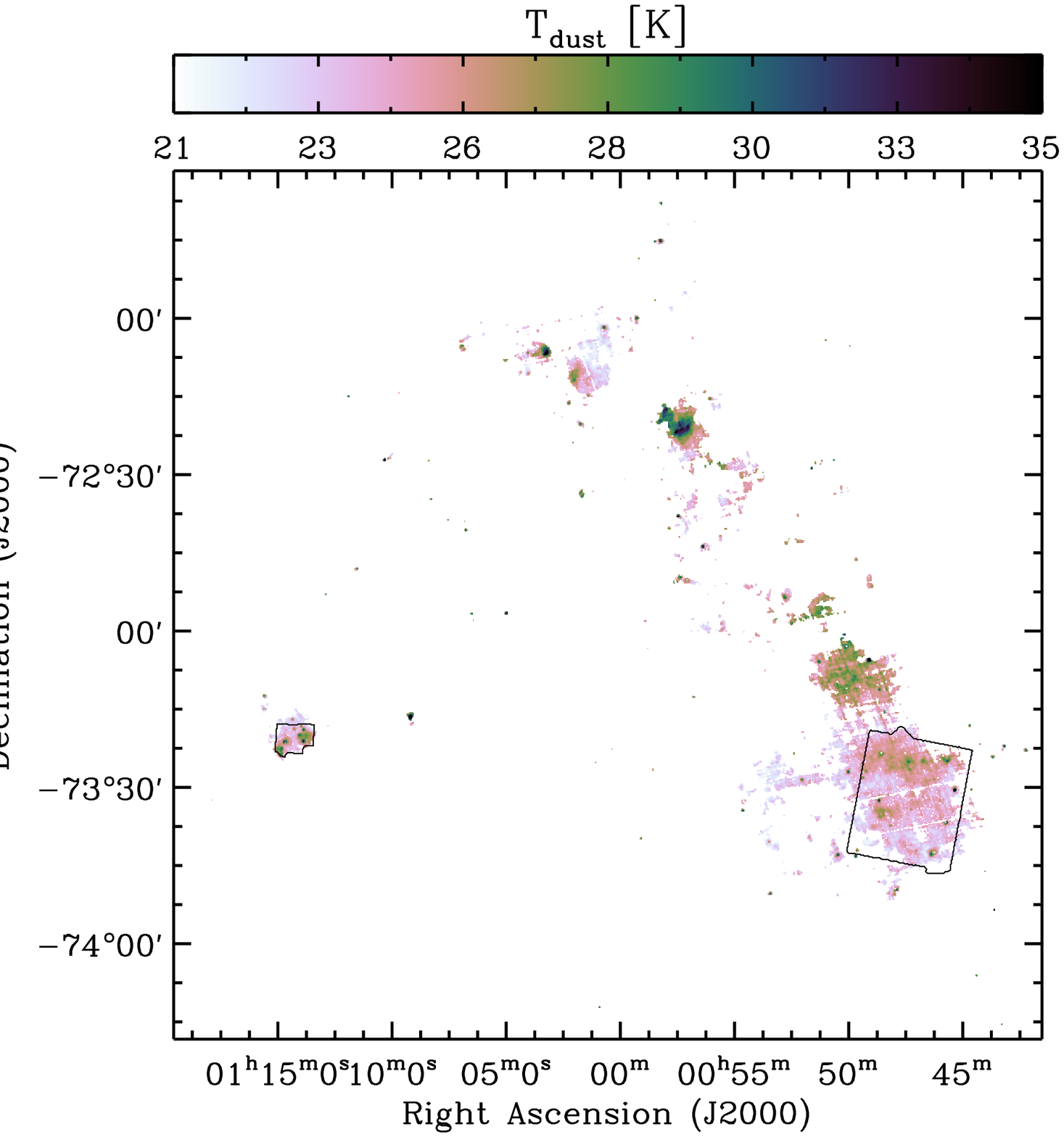} \includegraphics[width=0.49\linewidth]{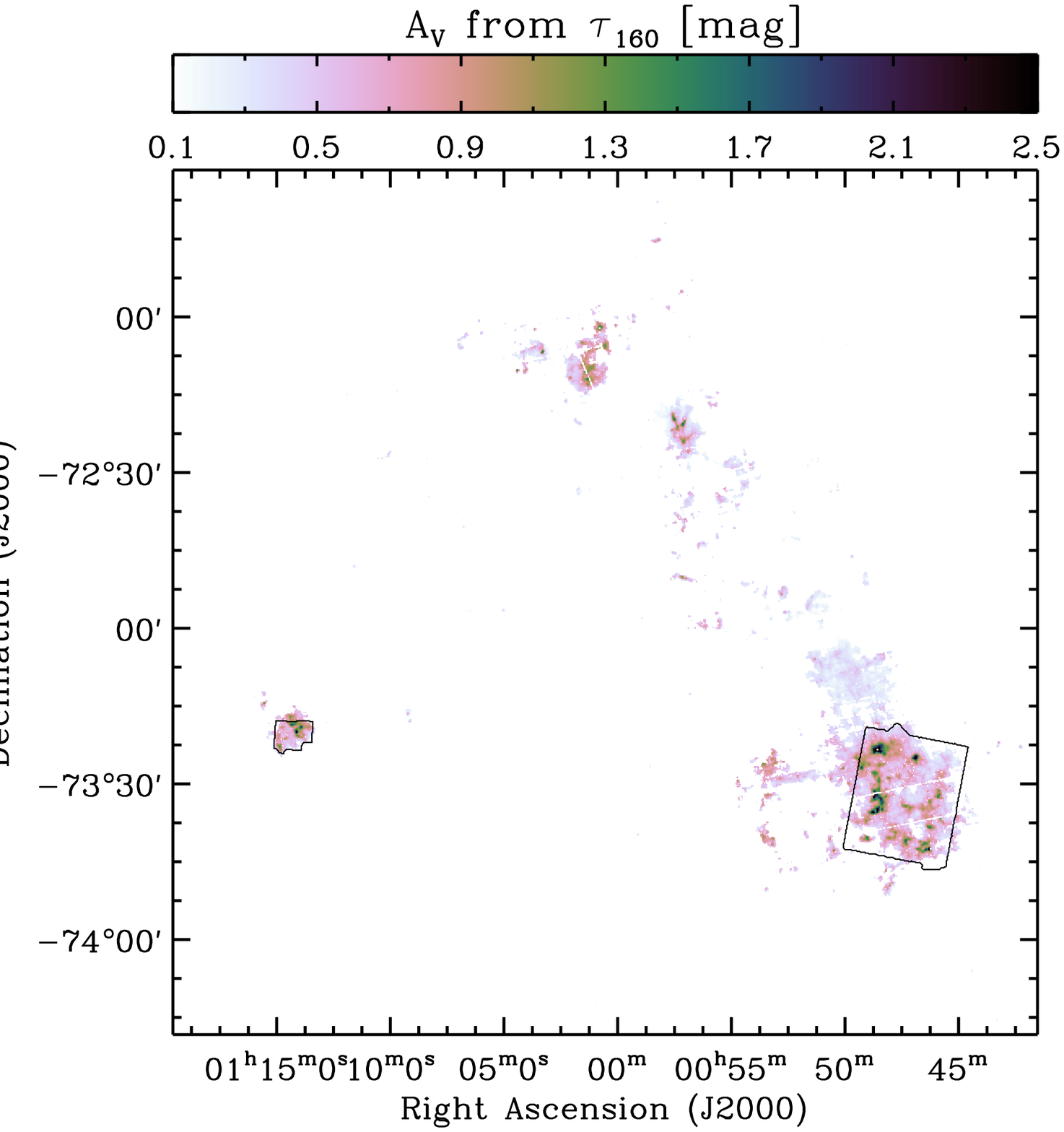} 
\caption{\label{fig:av_tdust_co_contour} Dust temperature (left) and $A_V$ (right; from $\tau_{160}$) maps of the LMC (upper panels) and SMC (lower panels). The fields
covered by CO data are shown in black contours. In addition to the majority of bright CO, the MAGMA field (black contours in the LMC) also shows enhancement of visual extinction 
relative to other regions in the LMC. That is, there is already a ``by-eye'' $I_{\rm CO}$-$A_V$ relation apparent in the figure.
(A colour version of this figure is available in the online journal.)} 
\end{figure*}

\subsubsection{Limitations of Our Approach}
\label{sec:limitation}

Our approach to estimate ``$A_V$'' has limitations, both due to our adopted approach and the use of IR emission
to trace dust. We model a single population of isothermal dust along each line of sight. In reality, a mixture of
temperatures and grain properties are present along each line of sight. This leads to biases in the total dust column determination 
({\rm e.g.,} see \citealt{2005ApJ...634..442S, 2006ApJ...640L..47S,2007ApJ...657..838S, 2008ApJ...684.1228S} for detailed
discussion) and could potentially affecting $A_{V}/\tau_{160}$. These  biases are somewhat alleviated by the inclusion of
the long wavelength {\em Herschel} data and the very high (by extragalactic standards) spatial resolution of {\em Herschel} at the LMC. 
Ultimately, they correspond to fundamental degeneracies in modelling the IR SED to derive a dust column. Resolution
clearly represents another limitation; while 10~pc resolution is the best achievable outside the Milky Way, this is still very coarse
compared to substructure observed in Milky Way clouds, so that measured ``$A_V$'' corresponds to something more like an
average extinction across a Milky Way cloud than a value within a cloud. Finally, the properties of the dust are expected to change
at some level, so that $A_V/\tau_{160}$ is not only uncertain in the absolute sense but may vary from location to location, e.g., due
to grain coagulation or the growth of mantles. This may produce some of the scatter in our comparison to direct extinction measures
above. Future observations with ALMA, HST, and ground based photometry all offer the potential to improve this work substantially. In the meantime, we present a first-order comparison using the best 
publicly available data.

\begin{table}
\begin{center}
\caption{Comparison of $T_{dust}$ and $A_{V}$ in MAGMA field}
\label{tab:spitzer_herschel}
\begin{tabular}{@{}lcccc@{}}
\hline\hline
& $<A_{V}>$ & $\sigma(A_{V})$ & $<T_{dust}>$ & $\sigma(T_{dust})$ \\ 
& (mag) & (mag) & (K) & (K) \\
\hline
Spitzer & 0.911 & 0.940 & 21.8 & 1.84\\
Herschel & 0.637 & 0.627 & 24.1 & 2.82\\
\hline\hline
\end{tabular}
\end{center}
\end{table}

\subsubsection{Comparison Between {\em Herschel} and {\em Spitzer} Results}
\label{sec:spitzer_herschel}

Before longer wavelength \emph{Herschel} data were available, we modelled IR emission in the LMC using \emph{Spitzer} 70 and 160~$\mu$m maps from SAGE survey \citep[][]{MEIXNER06}.
We used comparisons to coarser resolution data at 100$\mu$m to help us account for the out-of-equilibrium emission from 
very small grains (VSGs) contributing to the IR emission at 70~$\mu$m, finding about 50\% of the emission to represent contamination
but otherwise the approach was very similar to our main results here. Table~\ref{tab:spitzer_herschel} compares the median and standard 
deviation of $T_{dust}$ and $A_{V}$ in the MAGMA field between \emph{Spitzer} and \emph{Herschel}. On average, $T_{dust}$ from \emph{Spitzer} is $\sim 11$\% 
lower than our best-fit value from {\em Herschel}, while we find a $\sim 32\%$ higher $A_V$ using {\em Spitzer} than we do with {\em Herschel}. The point-by-point
correlation between the maps is good, with a Pearson coefficient of $p = 0.76$ and $0.9$ from comparing $T_{dust}$ and $A_V$ between maps
derived from the two telescopes. Consequently, the qualitative results of this paper would remain unchanged if we use dust properties derived from either 
telescope. For our purposes, the main change would be that the slope in the $I_{\rm CO}$-$A_V$ relation is somewhat lower if we use only {\em Spitzer} data.

As a sanity check, we also compare to \citet{SKIBBA12}, who used the HERITAGE data to derive dust temperature and dust mass in the Magellanic Clouds.
They fit a modified black body with $\beta = 1.5$ at $\lambda \le$ 300 $\mu$m  and a different $\beta$ that is allowed to vary to best describe the observation at 350 and 500 $\mu$m. 
We relate our estimate of $\tau_{160}$ to their dust mass estimate using their adopted mass absorption coefficient, $\kappa_{160} \approx 13.75~\rm cm^2g^{-1}$. Using this
$\kappa_{160}$, $D_{\rm LMC} = 50$~kpc, and $D_{\rm SMC} = 60$~kpc, our LMC $\tau_{160}$ map contains $\sim 1.0\times10^6 M_\odot$, within 10\% of the 
$\sim 1.1\times10^6 M_\odot$ found by \citet{SKIBBA12} and also similar to the $\approx 1.2 \times 10^6$~M$_\odot$ found using {\em Spitzer} data by \citet{BERNARD08}. In 
the SMC field, we find $\sim 0.9 \times10^5 M_\odot$ in our bright lines of sight, again just slightly below the $\sim 1.1\times10^5 M_\odot$ from \citet{SKIBBA12}. Considering 
potential issues with aperture matching and subtleties of fitting, our maps appear very consistent with previous works.

\subsection{Coverage and $A_V$ Distribution in the Magellanic Clouds}
\label{sec:completeness}

\begin{figure}
\includegraphics[width=0.99\columnwidth]{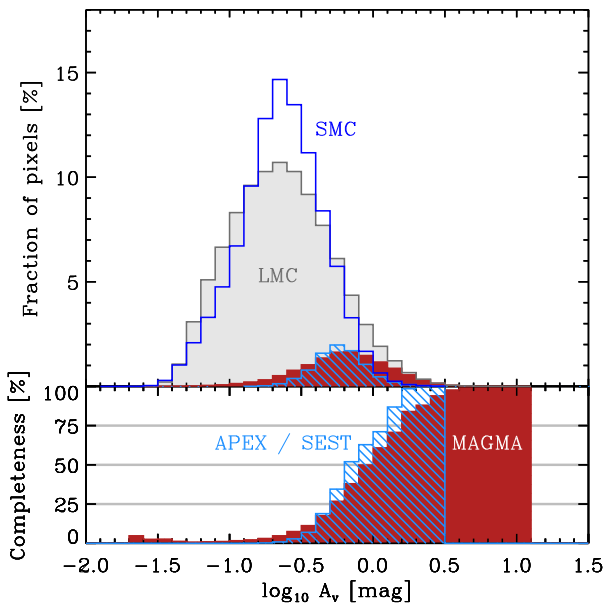}
\caption{\label{fig:comp} 
Top: Histogram of $A_V$ distribution in the LMC field (grey filled histogram; it includes all regions in the LMC with the {\em Herschel} IR intensities 
greater than their 3-$\sigma$ uncertainties) and the MAGMA field (red filled histogram; this is the region in the LMC field where the MAGMA survey mapped in CO). 
Likewise, the unfilled blue histogram shows distribution of $A_V$ in the SMC field, and  histogram shaded with blue diagonal lines shows that in the CO mapped regions in the SMC (the APEX and SEST fields).
Low extinction ($A_V \leq 1$~mag) lines of sight dominate the Magellanic Clouds and also comprise a large portion of the CO mapped area in each galaxy. 
Bottom: Completeness of $A_V$ in the CO mapped regions in the Magellanic Clouds (shown as red filled histogram for the MAGMA field, and histogram shaded with blue diagonal lines for the APEX and SEST fields), 
calculated by dividing the number of pixels with a given $A_V$
in the CO mapped region by the total number of pixels with that $A_V$ in each galaxy. The completeness level 
drops to $\approx 50$~per cent at $A_V \approx 0.8$~mag in the MAGMA field and at $A_V \approx 0.6$~mag in the APEX and SEST fields.
(A colour version of this figure is available in the online journal.)}
\end{figure}

As a targeted follow-up survey, MAGMA does not cleanly sample the distribution of $A_V$ in the LMC. Instead, MAGMA preferentially samples
high extinction lines of sight. Likewise, a similar bias in $A_V$ is expected for the APEX and SEST fields in the SMC.
Figure \ref{fig:av_tdust_co_contour} shows the maps of dust temperature ($T_{dust}$) and dust optical depth ($\tau_{160}$) scaled to visual extinction ($A_V$) in the LMC (upper panels) and SMC (lower panels) fields. 
The solid black contours show the regions covered by CO maps in each galaxy, the MAGMA field for the LMC and the APEX and SEST fields for the SMC. 
The upper-right panel clearly shows that dust shielding (estimated from $\tau_{160}$) in the MAGMA field is enhanced compared to other regions in the LMC.
This also appears to be the case for the APEX and SEST fields in the SMC. 
Considering the fact that the MAGMA field harbors the brightest molecular clouds identified by previous CO surveys in the LMC \citep{FUKUI99, FUKUI08},
even this simple visual comparison implies a close relation between $A_{V}$ and $I_{\rm CO}$ in the Magellanic Clouds.
In Figure~\ref{fig:av_tdust_co_contour}, we also note a quite significant variation of dust temperature across the Magellanic Clouds.
This reflects the variation of interstellar radiation field (ISRF) strength, which may impact the amount of CO emission 
in the region via photodissociation (e.g. \citealt{ISRAEL97}). We further explore this idea in Section~\ref{sec:tdust}, where we divide the MAGMA field into high and low $T_{\rm dust}$ regions.

In Figure~\ref{fig:comp}, we show histograms of $A_V$ distribution over the whole IR-bright area
and specifically over the CO mapped regions in the LMC and SMC. The bottom panel shows the completeness of $A_V$ coverage by the CO map in each galaxy, that is, the fraction of LMC and SMC pixels
in the specified $A_V$ bin (binsize of 0.1 dex mag) that lie within each galaxy's CO map. Therefore, 100\% in the bottom panel means that all pixels in that $A_V$ bin lie within CO map's field of view.

Low extinction ($A_V \leq 1$~mag) lines of sight make up most of the area in the Magellanic Clouds, even within the CO mapped regions where we observed enhanced dust shielding
relative to other regions of the LMC and SMC in Figure~\ref{fig:av_tdust_co_contour}.
On $\sim 10$~pc scale, the distribution of $A_V$ in the MAGMA field is well described by a log-normal function with mean 0.65~mag and standard deviation 0.30 dex.
The APEX and SEST fields in the SMC shows a narrower $A_V$ distribution, which can be fit by a log-normal function with mean 0.58~mag and standard deviation 0.17 dex. 
This is in qualitative agreement with the observations of local molecular clouds on sub-pc scale, where the column density distribution is well fit 
by a log-normal function, often accompanied by a power law tail \citep{KAIN09}.  At matched spatial resolution, we will see that the average extinction through a local 
Milky Way cloud is $\approx 1-2$~mag (Section~\ref{sec:int_cloudsdata}), a few times higher than the mean $A_V$ we see in the CO surveyed regions in the Magellanic Clouds. The $A_V$ associated with a 
Milky Way cloud would thus represent a bright spot, but not a dramatic outlier, in the Magellanic Clouds on $\sim 10$~pc scale.
 
The completeness of the MAGMA coverage exceeds 50~per cent at $A_V \sim 0.8$~mag. That is, about one half of the pixels in the LMC field with $A_V \sim 0.8$~mag 
lie within the MAGMA field. In the most extreme case where MAGMA recovers all of the CO emission from LMC then the bias in our $I_{\rm CO}$ estimate at a given $A_V$ will simply be the 
completeness in that $A_V$ range. So above $\approx 0.8$~mag, completeness introduces no more than a factor of two uncertainty. In reality, MAGMA does not recover all CO emission from
the LMC and we do not expect the measurement in the MAGMA field to be quite so strongly biased. In the SMC, the completeness of the APEX and SEST coverage becomes 50~per cent at lower $A_V$ than MAGMA, at $A_V \sim 0.6$, which is expected since the SMC CO map covers a large contiguous area in the southwest of the SMC.

\subsection{Milky Way Comparison Data}
\label{sec:mwdata}

We compare the $I_{\rm CO}$-$A_V$ relation in the Magellanic Clouds to the Milky Way using three data sets:
(1) analytic approximations to the highly resolved (sub-pc) $I_{\rm CO}$-$A_V$ relation measured in the Pipe nebula and Perseus molecular clouds,
(2) observations of local molecular clouds degraded to $\sim$ 10~pc resolution, and (3) the pixel-by-pixel relation for high Galactic latitude lines of sight
also convolved to $\sim$ 10~pc resolution.

\subsubsection{Highly Resolved Milky Way Clouds}
\label{sec:highresdata}

\begin{table}
\begin{center}
\caption{$I_{\rm CO}$-$A_V$ relation fitting parameters in the Pipe nebula and the Perseus molecular cloud}
\label{tab:highres}
\begin{tabular}{@{}lccccc@{}}
\hline\hline
Cloud & $I_0$ & $k$ & $A_{k12}$ & $A_{V}^{mid}$ & $b$ \\
& (\Kkmpers) & (mag$^{-1}$) & (mag) & (mag) \\
\hline
Pipe$^{a}$ & 32.3 & 0.694 & n/a & 4.55 & 0.036\\ 
Perseus$^{b}$ & 42.3 & 0.367 & 0.580 & n/a & n/a \\ 
\hline\hline
\end{tabular}
\end{center}
$^{a}$ $I_{\rm CO} =  I_{0}((1+e^{-k(A_{V}-A_{V}^{mid})})^{-1} - b)$ : \citealt{LOMBARDI06}

$^{b}$ $I_{\rm CO} =  I_{0}(1-e^{-k(A_{V}-A_{k12})})$ : \citealt{PINEDA08}
\end{table}

The proximity of Milky Way molecular clouds allows highly resolved 
comparisons of $I_{\rm CO}$ and $A_V$, with the limiting reagent mostly
wide field CO maps. We are aware of two quantitative studies of the dependence
of $I_{\rm CO}$ on $A_V$ in nearby clouds: \citet{LOMBARDI06} considered
the Pipe nebula and \citet{PINEDA08} studied Perseus. \citet{PINEDA10}
carry out a similar study of Taurus but do not analyze the relation in exactly the 
way we need for this comparison. These studies find
 
\begin{equation}
\label{eq:avco}
I_{\rm CO} = \left\{ \begin{array}{rl}
 I_{0}((1+e^{-k(A_{V}-A_{V}^{mid})})^{-1} - b) &\mbox{: Pipe} \\
 I_{0}(1-e^{-k(A_{V}-A_{k12})}) &\mbox{: Perseus}
       \end{array} \right.
\end{equation} 

\noindent where for Perseus $I_{0}$ is the integrated intensity at saturation, $A_{k12}$ is the
 minimum extinction needed to get CO emission, and $k$ is the conversion factor between the amount of 
 extinction and the optical depth. In the Pipe, the relation looks similar, but here $I_{0}(1-b)$ is the 
 saturation intensity, and the minimum extinction required for CO emission is equals to $A_V \sim A_V^{mid} - \frac{1}{k}\ln{\frac{1-b}{b}}$.
 We list the best fit parameters\footnote{Note that the best fit parameters for the Pipe nebula give negative
minimum extinction threshold, which means that $I_{\rm CO}$ is positive at $A_V = 0$.
These authors also fit $I_{\rm CO}-A_V$ relations for these clouds using a linear function (i.e. with the functional form 
$I_{\rm CO} = A_{V0} + rA_V$, where $A_{V0}$ is the minimum extinction threshold for CO emission and $r$ is the linear coefficient relating $A_V$ and $I_{\rm CO}$), and in this case
the minimum extinction threshold for CO emission in the Pipe nebula is positive.}
 reported in the above studies for these clouds in Table~\ref{tab:highres} and plot the two relationships as a point
 of comparison throughout the paper.

The qualitative behaviour of the relations observed in the Pipe and Perseus highlight some of the 
key physics governing CO emission from molecular clouds \citep[{\rm e.g.,} classic PDR models such as][]{MB88, VB88}. 
First, there appears to be a minimum amount of dust extinction required for bright CO emission. Below
this level, CO abundance is very low because of photodissociation, leading to no or negligible CO emission. 
Above that threshold there is an 
approximately linear relation between $A_V$ and $I_{\rm CO}$ for some range of $A_V$. Then at very high $A_V$, CO intensity 
saturates as the line becomes very optically thick across the whole velocity range. In this optically thick regime 
the observed CO intensity becomes a product of excitation temperature ($T_{ex}$), the beam filling factor, and linewidth. 
While this observed dependence of $I_{\rm CO}$ on $A_V$ highlight 
the importance of dust shielding for CO emission, the differences among the relation for the two clouds and even within an
individual cloud make it clear that $A_V$ is not the only factor that determines the amount of CO emission.
Different geometries and environmental factors (external radiation field, density structure, internal heating) will lead to a substantial
dispersion in CO emission even for the same amount of shielding. Indeed, one of the main conclusions of \citet{PINEDA08} was that environmental 
effects can be very strong even within the same molecular cloud complex (e.g., see their Figure~6).

\subsubsection{Integrated Measurements for Milky Way Clouds}
\label{sec:int_cloudsdata}

\begin{table*}
\begin{center}
\caption{Median $A_V$ and $I_{\rm CO}$ of Galactic Molecular Clouds at 3$^{\circ}$ resolution}
\label{tab:lores}
\begin{tabular}{@{}lccccccc@{}}
\hline\hline
Cloud & $l^{a}$ & $b^{a}$ & Area$^{a}$ & Pixels & $<A_V>^{b}$ & $<I_{\rm CO}>^{c}$ & $<I_{\rm CO} / A_V >^{b,c}$\\
& (degrees) & (degrees) & (deg$^2$) & & (mag) & (\Kkmpers) & (\Kkmpers~mag$^{-1}$) \\
\hline
Aquila & 20 & 8 & 227 & 85 & 4.0 & 7.5 & 2.0\\
Camelopardalis & 148 & 20 & 159 & 61 & 0.54 & 0.32 & 0.55\\
Chamaeleon & 300 & -16 & 27 & 7 & 1.4 & 2.0 & 1.6\\
Gum Nebula & 266 & -10 & 97 & 37 & 1.5 & 0.46 & 0.31\\
Ophiuchus & 355 & 17 & 422 & 151 & 1.9 & 2.6 & 1.3\\
Orion & 212 & -9 & 443 & 163 & 2.1 & 1.6  & 0.65\\
Polaris Flare & 123 & 24 & 134 & 55 & 0.81 & 2.5 & 2.7 \\
Taurus & 170 & -15 & 883 & 313 & 1.9 & 3.5 & 1.8 \\
\hline\hline
\end{tabular}
\end{center}
$^{a}$ {Compiled from Table~1 in \citet{DAME01}.}

$^{b}$ {Computed from {\em Planck} $E(B-V)$ map, assuming $R_V$ = 3.1. Note that this $A_V$ may be systematically overestimated (see text).}

$^{c}$ {Computed from \citet{DAME01} CO map.}
\end{table*}

To make a more direct comparison of the Milky Way to the Magellanic Clouds,  we consider local Galactic molecular clouds from Table~1 in \citet{DAME01}.
We calculate their average $A_{V}$ and $I_{\rm CO}$ values on $\sim 10$~pc scales to simulate how they might appear in one of our Magellanic Cloud maps
and report the measurements in Table~\ref{tab:lores}. We compile the $A_{V}$ and $I_{\rm CO}$ values of these clouds from the Milky Way CO map by 
\citet{DAME01} and the $E(B-V)$ map published by the {\em Planck} collaboration \citep{PLANCK13_DUST}, where we take $R_V = 3.1$ to convert $E(B-V)$ into $A_V$. We refer the interested reader to 
\citet{PLANCK13_DUST} for more detailed information on the {\em Planck} $E(B-V)$ map and note that $A_V$ in local molecular clouds calculated from the 
{\em Planck} $E(B-V)$ map may be slightly overestimated
\footnote{The {\em Planck} $E(B-V)$ map used in this analysis is a conversion of their dust optical depth ($\tau$)
to $E(B-V)$ using the correlation between $E(B-V)$ of SDSS quasars and dust optical depth at high Galactic latitude lines of sight ({\rm i.e.,} the conversion we 
discussed in Section~\ref{sec:tau_to_av}, see Figure~22 in \citealt{PLANCK13_DUST}). However, \citet{PLANCK11_MC} compared column density $N(\rm H)$ from 
NIR extinction with dust optical depth, finding higher dust emissivity, $\tau/N(\rm H)$, in the molecular phase by a factor $\sim 2$ than the atomic phase (see also Figure~20 in \citealt{PLANCK13_DUST}). 
Therefore, we caution that the {\em Planck} $E(B-V)$ map may systematically overestimate the actual $E(B-V)$ (and $A_V$) in local clouds. 
For example, comparison of $E(B-V)$ from NIR extinction and the {\em Planck} $E(B-V)$ map in Taurus and Ophiuchus 
molecular clouds \citep{PLANCK13_DUST} show that correlations between them are pretty strong (their Table~5), but the {\em Planck} $E(B-V)$ is systematically higher by 25 per cent.}.

To simulate the $\sim 10$~pc physical resolution of our Magellanic Cloud data, we convolve the {\em Planck} $A_V$ map (original resolution 5\arcmin) and the \citet{DAME01} CO map
(original resolution 7.5\arcmin) with a 3$^{\circ}$ Gaussian kernel. At $200$~pc, a typical distance to a nearby molecular cloud, this resolution corresponds to $\approx 10.5$~pc. 
Most local clouds subtend several pixels at this resolution (7 pixels for the Chamaeleon molecular cloud at the least; at the most 313 pixels for the Taurus-Perseus-Auriga complex). 
We take the median $A_{V}$ and $I_{\rm CO}$ across all of these pixels as representative and report them in Table~\ref{tab:lores}; we adopt the $A_V$ and $I_{\rm CO}$ of each cloud 
at $\pm$ 1-$\sigma$ percentiles as the representative uncertainties when plotting the data.

\subsubsection{High Galactic Latitude ($|b| > 5^{\circ}$) Emission}
\label{sec:highlatdata}

As a final point of comparison, we use all-sky Milky Way CO and $A_V$ maps to explore the high Galactic latitude lines of sight.
We use the same {\em Planck} $E(B-V)$ map\footnote{Unlike the case for the local clouds, high Galactic latitude lines of sight
are diffuse and the application of $E(B-V)/\tau$ conversion is not expected to overestimate the actual $E(B-V)$ value in this case.} 
described above section to calculate $A_V$. Because the \citet{DAME01} CO map only covers limited range of Galactic latitudes ($|b| < 30^{\circ}$),
we use the the {\em Planck} TYPE 2 CO map. This is a map of integrated CO line emission that has been extracted from the {\em Planck} HFI channels
using the multi-channel method by the {\em Planck} team \citep{PLANCK13_CO}. This map has an angular resolution of 15\arcmin, and is known to be better 
suited for intermediate / high Galactic latitude regions than the TYPE 1 CO map \citep{PLANCK13_CO}. 

Before comparing this map with the $A_V$ map, we corrected for the contribution of $^{13}$CO to the map by dividing the TYPE 2 CO map by 1.2.
We then degraded the resolution of the CO map to match that of the $A_V$ map. Using the $A_V$ and CO maps described above, we construct a 
pixel-by-pixel $I_{\rm CO}$-$A_V$ relation for the high latitude Milky Way sky suitable for comparison to Magellanic Cloud measurements.
We consider all area with $|b| > 5^{\circ}$, avoiding the Milky Way midplane in order to remove confusion from multiple components along a line of sight. 

\begin{figure*}
\includegraphics[width=0.49\linewidth]{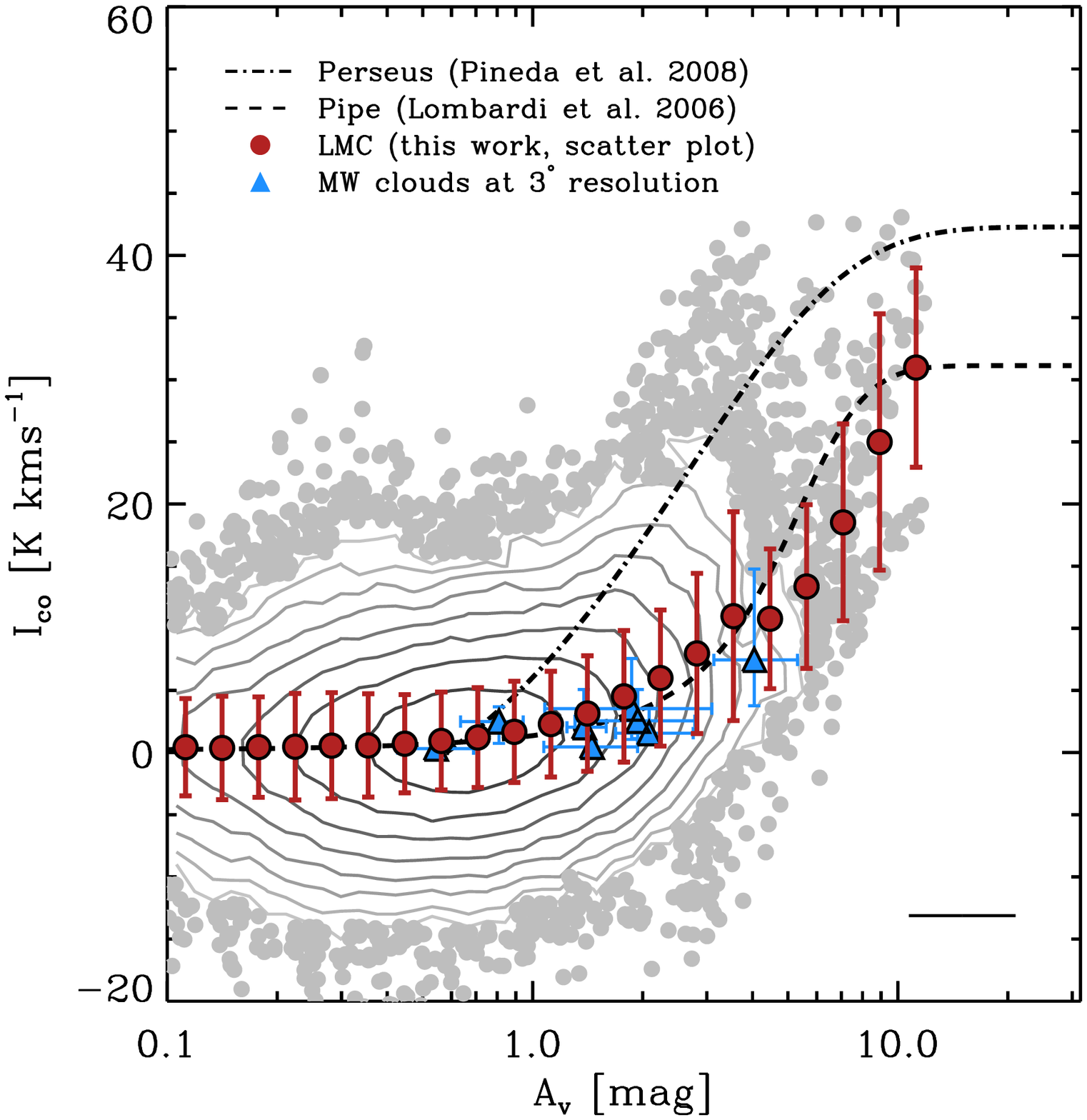} \includegraphics[width=0.49\linewidth]{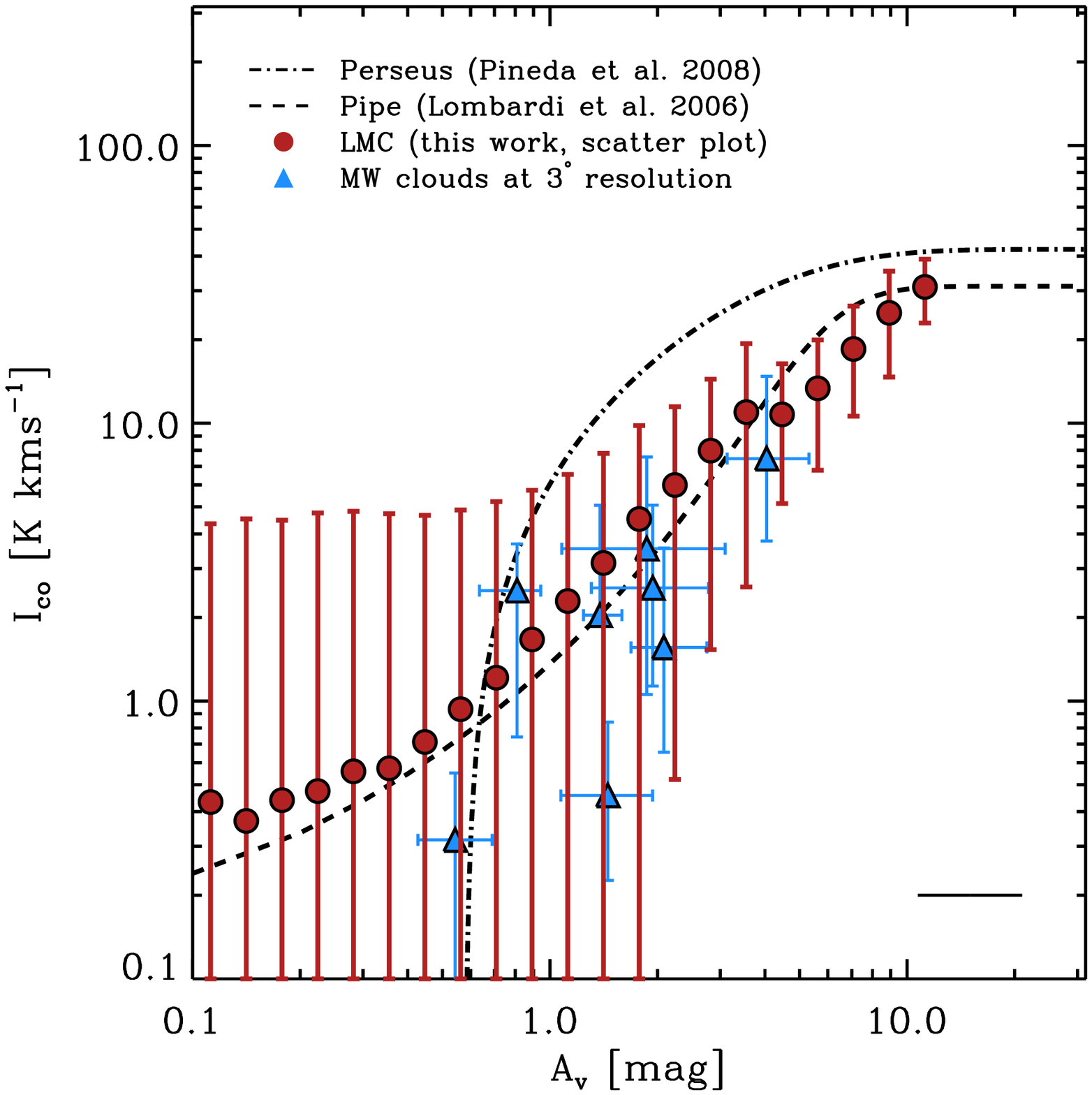} 
\includegraphics[width=0.49\linewidth]{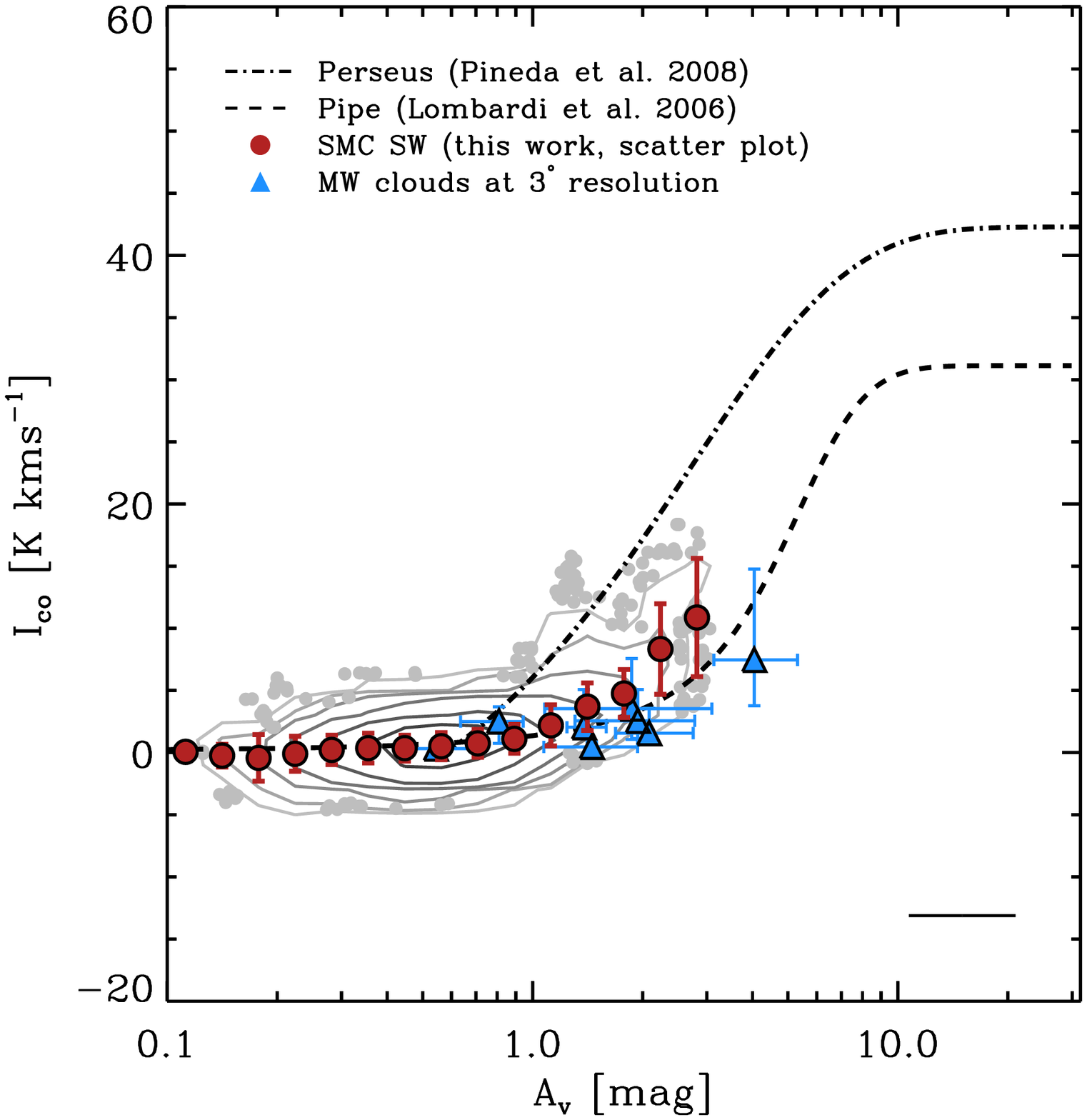} \includegraphics[width=0.49\linewidth]{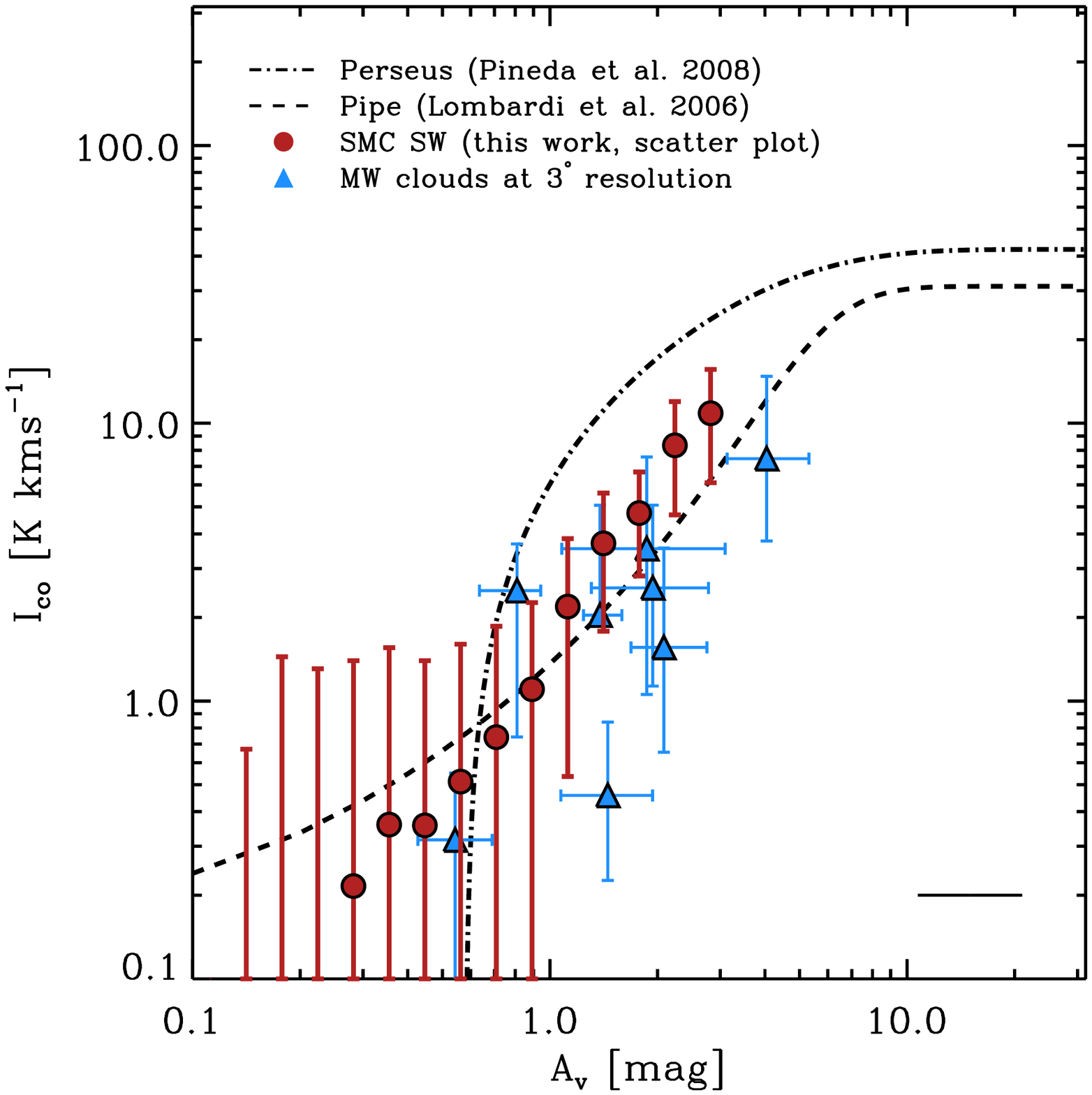} 
\caption{\label{fig:avco} Pixel-by-pixel comparison of $A_{V}$ ($x$-axis) and $I_{\rm CO}$ ($y$-axis) in the LMC (top panels) and SMC (bottom panels),
shown in semi-log scale in the left panels and in log scale in the right panels. 
In the left panels, the distribution of data points in the LMC and SMC are shown as the contours, while the grey points show individual 
lines of sight where the density falls below 10 points per bin. 
The lowest contour means there are at least 10 individual data points in the bin, increasing by factors 
of 2 with each step for the LMC and by factors of 3 for the SMC. 
The red circles with error bars indicate the median CO intensity and its scatter measured in bins of log $A_{V}$,
our LMC and SMC $I_{\rm CO}$-$A_V$ relations. 
For comparison, we also plot the $I_{\rm CO}$-$A_V$ relations measured at much higher ($\sim$ sub-pc) resolution for two Galactic
clouds, Perseus (dash-dotted, \citealt{PINEDA08}) and the Pipe (dashed : \citealt{LOMBARDI06}). 
Blue triangles represent a more well-matched Milky Way comparison, Galactic molecular cloud data compiled from {\em Planck} 
\citep{PLANCK13_DUST} and \citet{DAME01}, degraded to physical resolution similar to what we achieve in the Magellanic Clouds 
($3^\circ$, which corresponds to $\sim 10.5$ pc at the distance of 200 pc). 
In the semi-log plots one can see how median CO intensity approach zero as $A_V$ decreases in the Magellanic Clouds, while
in the log-log plots we rescale the $y$-axis to better compare $I_{\rm CO}$-$A_V$ relations in the Magellanic Clouds with Galactic molecular clouds. 
The horizontal error bar in the bottom-right corner of each panel shows a typical uncertainty in $A_V$ derived from Monte-Carlo analysis (Section~\ref{sec:unc}).
(A colour version of this figure is available in the online journal.)}
\end{figure*}

\begin{figure}
\includegraphics[width=0.99\columnwidth]{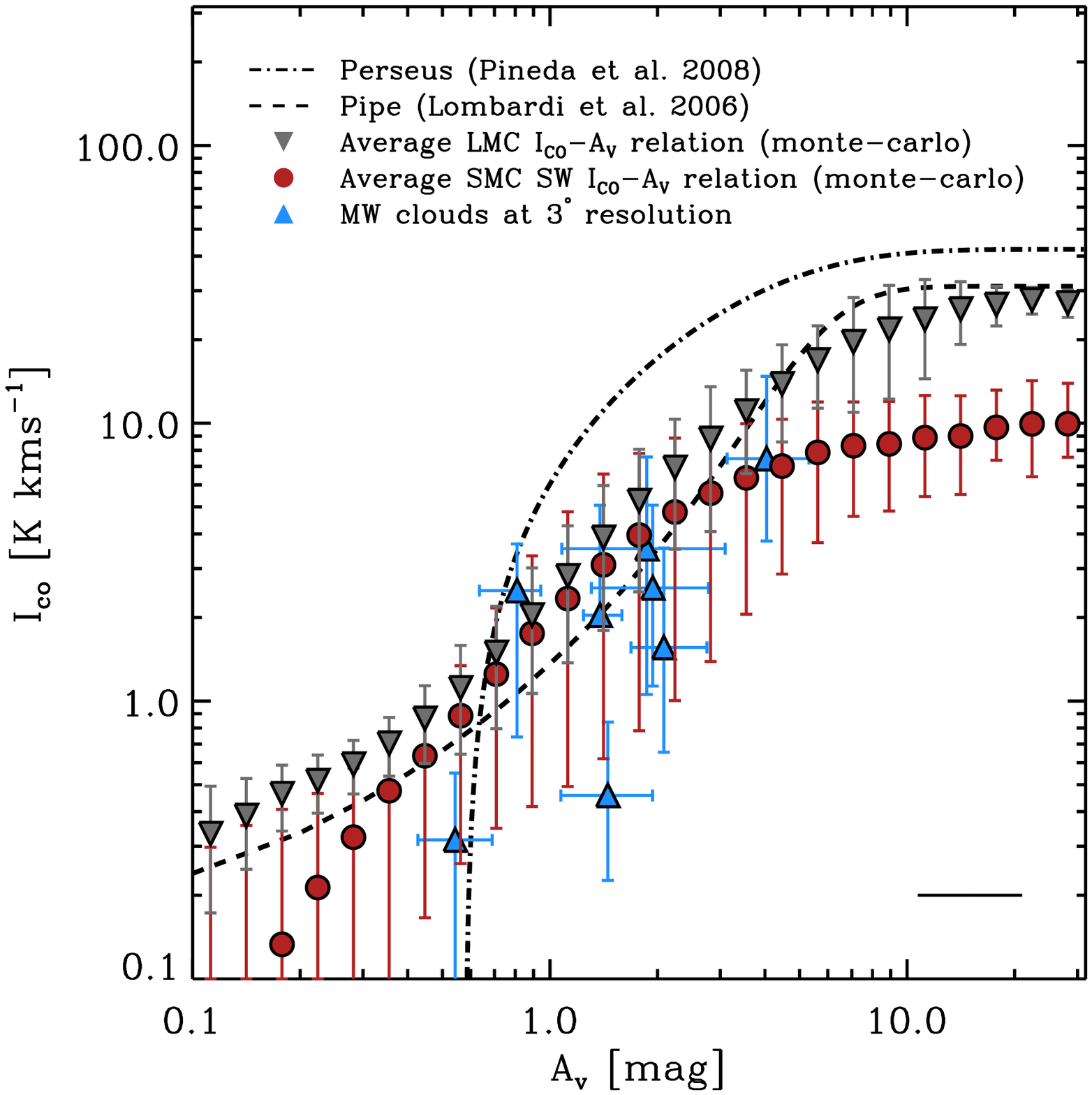}
\caption{\label{fig:avco_monte} 
Average $I_{\rm CO}$-$A_V$ relation in the LMC (grey downward triangles) and SMC (red circles) from Monte-Carlo simulation (Section {\ref{sec:unc}}). 
Other lines and symbols are the same as Figure~\ref{fig:avco}. Here, the vertical error bars associated with the average relations do represent 
uncertainty in the average relations. Note that high $A_V$ points in the average relations arise from very few lines of sight moving across $A_V$ bins as we simulate $A_V$ maps and thus mimic saturation effect observed in highly resolved Galactic molecular clouds (see Section~\ref{sec:highresdata}). The agreement between the LMC, SMC, and integrated measurements for the Milky Way is fairly good at $A_V \le 4$, suggesting that dust shielding is the primary factor that determines the distribution of bright CO emission from a molecular cloud.
(A colour version of this figure is available in the online journal.)}
\end{figure}

\section{Results}
\label{sec:results}

\subsection{$I_{\rm CO}$ vs. $A_{V}$ in the Magellanic Clouds}
\label{sec:mcavco}

Figures~\ref{fig:avco} and \ref{fig:avco_monte} show our main observational results, the $I_{\rm CO}$-$A_V$ relationship in the Magellanic Clouds.
For this analysis, we consider the MAGMA field for the LMC and the APEX field for the SMC. For the APEX CO data, we assume a ratio of unity
to translate CO $J = 2 \to 1$ to CO $J = 1 \to 0$ \citep[e.g., see][]{BOLATTO03}. For clarity, the $I_{\rm CO}$-$A_V$ relationship in the N83 complex is not displayed in 
Figures~\ref{fig:avco} and \ref{fig:avco_monte}, but we note that it is similar to the $I_{\rm CO}$-$A_V$ relationship in the southwest region of the SMC. 

Figure~\ref{fig:avco} shows the relationship between $I_{\rm CO}$ and $A_V$ pixel-by-pixel for each galaxy (the LMC in upper panels and the SMC in lower panels).
The distribution of individual data points are shown as the contours and grey points in the left panels (semi-log scale). 
The median and 1-$\sigma$ scatter of the data are shown as the red circles (binned by log $A_V$) in both the left (semi-log scale) and right (log scale) panels.
In the left panel of Figure~\ref{fig:avco}, we find that the $I_{\rm CO}$-$A_V$ relation for individual lines of sight shows large scatter, 
but the median trend (red circles) confirms the impression from simple visual comparison in Figure~\ref{fig:av_tdust_co_contour}. 
That is, the CO intensity appears to increase as a function of $A_V$ in the Magellanic Clouds.

In Figure~\ref{fig:avco_monte}, we plot this average relation for both galaxies. Here the 
mean and uncertainty derive from Monte-Carlo simulation (Section~\ref{sec:unc}), so that the error bars reflect the uncertainty in the
mean across 1,000 realizations for the LMC and an equivalent uncertainty for the SMC. 
These average relations in Figure~\ref{fig:avco_monte} are thus our best estimate of how CO intensity ($I_{\rm CO}$)
depends, on average, on dust shielding ($A_V$) at cloud-scales in the LMC and SMC. The estimate
takes into account the systematic uncertainties in estimating $A_V$ from the IR emission. Interpreting the error bars requires
some care; many of the factors that we simulate create correlated errors across a whole realization. Therefore one should
largely view the error bars as the space within which the mean relation could move.

Care must also be taken when interpreting the high $A_V$ portion of the average relations. Recall that low $A_V$ ($A_V \le 1$) dominate the Magellanic Clouds, and 
there are very few data points in the high $A_V$ bins in the LMC and SMC (e.g. see Figure~\ref{fig:comp}). 
For example, there are only 15 lines of sight that has $A_V$ greater than 10 in the LMC, and no line of sight has $A_V$ greater than 4 in the SMC.
The greater number of points at lower $A_V$ will lead those points to preferentially scatter to high $A_V$ and contaminate
the measurements at high $A_V$ in the Monte Carlo calculation. This will artificially lower the mean $I_{\rm CO}$ at high $A_V$ in the Monte 
Carlo simulation.

With this in mind, we note that except at $A_V \ge 4$ in Figure~\ref{fig:avco_monte}, the LMC and SMC in Figures~\ref{fig:avco} and \ref{fig:avco_monte}
overlap one another, showing similar CO intensity at a given $A_V$. This agreement between the LMC and SMC despite their factor of $\approx 2$ 
difference in metallicity is consistent with the theoretical picture that CO intensity depends on dust shielding ($A_V$) in an approximately universal way.
In the next Section, we will compare the Magellanic Clouds to the Milky Way in several ways to see that  this holds true from $Z \approx 1/5~Z_\odot$
up through solar metallicity.

We noted earlier that the measured $I_{\rm CO}$ scatters considerably at a given $A_{V}$ in Figure~\ref{fig:avco}. 
Much of this scatter reflects the noise in the CO map, and this scatter clearly dominates the distribution of individual data below $A_V \approx 1$~mag.
Still, we verified that a real CO signal emerges as we stack the CO spectra in the LMC in bins of $A_V$ (see Figure~\ref{fig:spec_tpk_lmc_fit}).
At higher extinctions, $A_V \geq 1$~mag, we also see substantial additional scatter about the median CO intensity in fixed $A_V$ bins. We interpret this as real astronomical signal,
indicating that line-of-sight $A_V$ on 10~pc scale is not a perfect predictor of $I_{\rm CO}$ even in the absence of noise.
This is reasonable and expected as many physical effects beyond shielding may influence CO emission, for example,
variations in cloud structure, geometry, chemistry, and the interstellar radiation field. 
The important caveat to bear in mind is that while $I_{\rm CO}$ does depend on $A_V$, \emph{the relationship emerges
only after substantial averaging} because individual lines of sight have large scatter in $I_{\rm CO}$-$A_V$ parameter space.

\subsection{Comparison to the Milky Way}
\label{sec:mw}

We are primarily interested in testing the hypothesis that CO emission depends on dust shielding ($A_V$) in the same manner
across environment that differ in metallicity. In the previous Section, we saw that CO intensity at a given $A_V$  is indeed similar in the LMC and SMC. 
In this Section, we compare our Magellanic Cloud results to the Milky Way in three different ways.

First, we compare the Magellanic Clouds to the highly resolved (sub-pc resolution) $I_{\rm CO}$-$A_V$
relation measured in the Pipe nebula and Perseus molecular clouds. These appear as the dashed line (Pipe) and the dash-dotted line (Perseus)
in Figures~\ref{fig:avco} and \ref{fig:avco_monte}. Overall, the $A_{V}$-$I_{\rm CO}$ relation in the Magellanic Clouds resemble that in the Pipe nebula
(note that the Pipe curve is not a fit to our data or even normalized to match our data), 
while the Magellanic Clouds exhibit $2$--$3$ times fainter CO emission than the Perseus molecular cloud at high $A_{V}$.
Most importantly, neither the LMC nor the SMC in Figure \ref{fig:avco} exhibit 
the qualitative features observed in the Pipe and Perseus (Section~\ref{sec:highresdata}), at least not prominently.
We observe no clear saturation in $I_{\rm CO}$ and the threshold behavior, if present, appears far weaker than 
in the Perseus molecular cloud. 
The average $I_{\rm CO}$-$A_V$ relations in Figure \ref{fig:avco_monte} seem to exhibit 
a saturation of CO at high $A_V$ ($A_V \ge 10$ mag for the LMC, $A_V \ge 4$ for the SMC), but as noted earlier, 
the apparent trends observed at high $A_V$ in the average $I_{\rm CO}$-$A_V$ relations are vulnerable to 
artifacts arising from Monte-Carlo simulation.

The lack of evidences for the saturation of CO and threshold behaviour in the Magellanic Clouds almost certainly stem from the dramatic difference in
resolution between our measurement in the Magellanic Clouds and those used to construct the Pipe and Perseus curves. 
Our measurements combine large parts of a cloud ($\sim 10$ pc) into a single data point, so that each line of sight is an average of 
$A_V$ and $I_{\rm CO}$ in a 10 pc beam. On the other hand, the Pipe and Perseus relations are measured at much higher resolution (sub-pc).
These clouds would not exhibit $A_V \sim 5$--$10$~mag at 10 pc resolution.
For a fairer comparison to these highly resolved curves, one would need very high resolution $A_V$ and $I_{\rm CO}$ data in the Magellanic Clouds.

An alternative approach is to measure $A_V$ and $I_{\rm CO}$ averaged over a 10~pc beam for the local clouds. These appear
as blue triangles in Figures~\ref{fig:avco} and \ref{fig:avco_monte}.  These triangles represent our best estimate of 
how local clouds would look like at the distance of the Magellanic Clouds. This makes them analogous to the individual grey points in Figure~\ref{fig:avco}.
At matched $\sim 10$~pc resolution, most of the Milky Way clouds have rather low $A_V$, ranging from $\sim$ 0.5~mag to $\sim$ 4.0~mag. 
The agreement between the Milky Way and the Magellanic Clouds is much better in the case of matched resolution than for the highly resolved Milky Way
relations. A close inspection of Figure~\ref{fig:avco_monte} suggests that some of the local clouds have rather 
fainter CO emission at a given $A_V$ than the average lines of sight in the Magellanic Clouds, but this could be partly due to overestimated $A_V$ in the local clouds
in the {\em Planck} maps (Section~\ref{sec:int_cloudsdata}). Given that the scatter in $I_{\rm CO}$ is very large at a given $A_V$ both for the Magellanic Clouds 
and the Milky Way clouds, we interpret the integrated measurements of $I_{\rm CO}$-$A_V$ in local clouds to substantially agree with our measurements in the
Magellanic Clouds. This reinforces our results from the internal comparison of the two Magellanic Clouds.

\begin{figure}
\includegraphics[width=0.99\columnwidth]{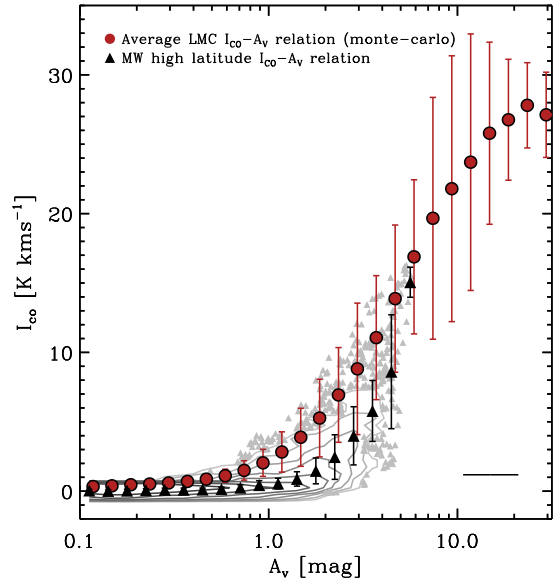}
\caption{\label{fig:mw_avco} Integrated CO intensity ($y$-axis) as a function of $A_V$ ($x$-axis) at high Galactic 
latitude ($|b| > 5^\circ$) lines of sight in the Milky Way, compiled from {\em Planck} data \citep{PLANCK13_DUST} degraded 
to $\sim 3^{\circ}$ resolution. The contours and grey points show the distribution of data points 
in the $I_{\rm CO}$-$A_V$ space, following the same contour definition for the LMC in the left panel of Figure~\ref{fig:avco}.
The median CO intensity at a given $A_V$ is shown as the black triangles, and the average LMC $I_{\rm CO}$-$A_V$ 
relation (Figure~\ref{fig:avco_monte}) is shown as the red circles, shifted slightly toward right in $x$-axis for easier comparison.
(A colour version of this figure is available in the online journal.)}
\end{figure}

As a final comparison, we plot the $I_{\rm CO}$-$A_V$ relation for high latitude ($|b| > 5^{\circ}$) Milky Way lines of sight in Figure~\ref{fig:mw_avco}.
The high latitude Milky Way lines of sight (black triangles) exhibit a similar shape to the Magellanic Cloud relations but with a tendency 
towards lower $I_{\rm CO}$ at a given $A_V$.  That is, the Magellanic Clouds appear to be brighter than the high latitude Milky Way 
in CO on average for a given $A_V$. Quantitatively, the median $I_{\rm CO}$/$A_V$ for Milky Way high latitude lines of sight is
$\approx 0.5$~K~km~s$^{-1}$~(mag)$^{-1}$, which is about 0.25 times the value for the LMC ($\approx 1.9$~K~km~s$^{-1}$~(mag)$^{-1}$).

Our best explanation for this difference is that the Milky Way values are likely to be biased low by 
dust associated with a  long path length through the Milky Way H{\sc i} disk. For example, at $b \sim 10^{\circ}$ a 200~pc thick H{\sc i} 
disk will yield an integrated path length of roughly a kpc. This path length will preferentially sample atomic gas, which 
has a higher scale height than molecular gas and little associated CO emission. More, the dust along that line of sight through and
extended disk will contribute little to shielding distant CO from dissociating radiation. Such effects will undermine any mapping between
line of sight extinction and the local dust shielding that should affect CO emission. In Appendix~\ref{sec:appendix_hi_corr}, we show that
a simple correction for dust associated with an extended H{\sc i} disk leads to a median $I_{\rm CO}$ at a given $A_V$ for the high latitude 
Milky Way that more closely resembles what we find in the Magellanic Clouds.

The Magellanic Cloud data may also be biased high by our focus on the CO mapped regions.
As shown in the bottom panel of Figure~\ref{fig:comp}, the completeness of $A_V$ in the MAGMA field
drops to $\sim 50$~per cent at $A_V \sim 0.8$~mag, and at $A_V \sim 0.6$~mag in the APEX and SEST fields. 
This means that at $A_V \sim 0.8$~mag, $I_{\rm CO}$/$A_V$ in the LMC has a lower limit of about half the current value
if we assume that the other half lines of sight not mapped in CO do not have associated CO emission at all, and
a similar logic can be applied to the SMC as well.
If this is the case, $I_{\rm CO}$ at a given $A_V$ would be closer between the Magellanic Clouds and high latitude Milky Way.
Even if this bias drives the results, Figure~\ref{fig:mw_avco} suggests the somewhat surprising result
that the active parts of the Magellanic Clouds are better at emitting CO than the high latitude Milky Way.

Overall, the sense of the comparison made in Section~\ref{sec:mw} is this: the Magellanic Clouds, at least the bright regions covered
by CO surveys, show $I_{\rm CO}$ at a given $A_V$ comparable to or somewhat below than those found at highly resolved (sub-pc) local Galactic clouds. 
The Magellanic Clouds data do not show clear evidence for saturation of CO line and $A_V$ threshold observed in high resolution Milky Way data, likely due to low resolution. 
After accounting for resolution differences,
we find that a sample of Milky Way clouds at 10~pc resolution largely overlaps the average $I_{\rm CO}$-$A_V$ relations in the Magellanic Clouds.
Taking a broader view and considering high latitude emission from the Milky Way, we find a qualitatively similar $I_{\rm CO}$-$A_V$
relation to the Magellanic Clouds but note important quantitative differences with the sense that at a fixed $A_V$, gas in the high latitude Milky Way
emits somewhat less CO than the regions covered by the CO maps in the Magellanic Clouds.

\subsection{Influence of the Interstellar Radiation Field}
\label{sec:tdust}

\begin{figure}
\includegraphics[width=0.99\columnwidth]{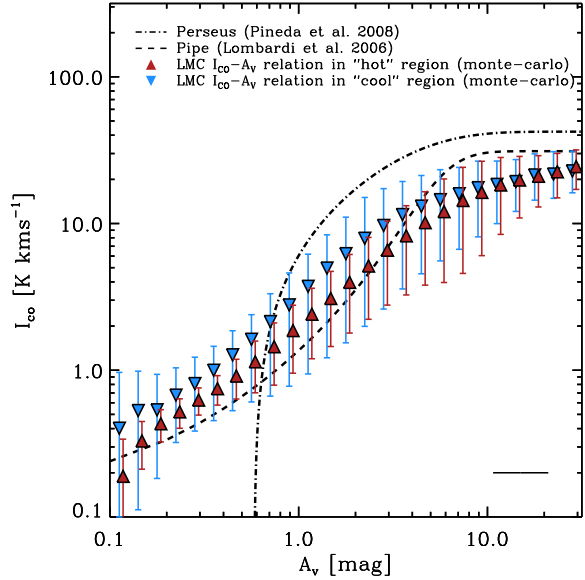}
\caption{\label{fig:tdust} Temperature dependence of LMC $I_{\rm CO}$-$A_V$ relation. 
We divide the MAGMA field into hot ($T_{dust} > 24.1$~K; the red upward triangles with error bars, shifted slightly toward right in $x$-axis for easier comparison) 
and cool  ($T_{dust} < 24.1$~K; the blue downward triangles with error bars) regions 
based on the median dust temperature ($\sim 24.1$~K). 
Interestingly, the region with hotter dust temperature (and more intense ISRF)
shows fainter CO emission measured at a fixed dust shielding ($A_V$) compared to the region with colder dust temperature.
This is the opposite of what one would expect for the region with hotter dust temperature having stronger CO emission
due to a higher kinetic temperature, and we interpret this as the evidence for an intense ISRF to suppress CO emission at 
a fixed dust shielding in the hot region more than the cool region.
(A colour version of this figure is available in the online journal.)}
\end{figure}

Figure \ref{fig:avco} showed large scatter in $I_{\rm CO}$ at fixed $A_V$, implying additional physics beyond
the abundance of dust at play.
One simple and often-discussed ``second parameter'' is the interstellar radiation
field \citep[ISRF][]{ISRAEL97, PINEDA09}. We have estimated this quantity as part of our dust modelling: the dust
temperature, $T_{dust}$, depends on the strength of ISRF such that $X_{ISRF} \sim T_{dust}^{4+\beta}$). The
radiation field drives the photodissociation of CO molecules, so that we may reasonably expect $T_{dust}$ to affect
$I_{\rm CO}$ within a bin of fixed $A_V$.

Figure \ref{fig:tdust} shows a simple test for the effect of $T_{dust}$ as a second parameter (as an extension to this test, we also discuss the effect of strong radiation field on the $I_{\rm CO}$-$A_V$ relation in Appendix~\ref{sec:appendix_30dor}, by focusing on 30 Doradus complex in the LMC). For this experiment, we divide our LMC data into
two bins: a ``cool'' region with lower-than-median $T_{dust}$, {\rm i.e.}, $T_{dust} < 24.1$~K, and a ''hot'' region 
with higher-than-median $T_{dust}$, $T_{dust} > 24.1$~K (See Figure~\ref{fig:av_tdust_co_contour} for the dust temperature map). 
The median dust temperatures of these two regions are $\sim 26.0$~K (``hot'')
and $\sim 22.7$~K (``cool''). With our fiducial $\beta = 1.5$, this temperature difference corresponds to difference in
ISRF strength of a factor of $\sim 2.1$. Note that the ISRF in the MAGMA field is already on average greater by a factor of $\sim 5.8$ than the Solar 
Neighborhood, where $T_{dust} \sim 17.5$~K. 
We calculate the average $I_{\rm CO}$-$A_V$ relations in the ``hot'' and ``cool'' regions and estimate their uncertainties using Monte-Carlo simulation.

Interestingly, Figure \ref{fig:tdust} shows that $I_{\rm CO}$ at fixed $A_V$ is higher in the cool region (blue downward triangles) than the hot region (red upward triangles) in the MAGMA field.
This observed trend supports the idea that an intense ISRF suppresses CO emission, since several effects might lead to 
the opposite direction of what we see in Figure \ref{fig:tdust}.
First, the uncertainties on $\tau_{160}$ and $T_{dust}$ are correlated in the sense that underestimating $T_{dust}$ would lead us to 
overestimate $\tau_{160}$. This has the opposite sense of what we see here and so seems unlikely to drive the separation in Figure
\ref{fig:tdust}. Second, if a higher ISRF leads to higher excitation for the CO, we might expect a higher brightness temperature in regions
with high $T_{dust}$. Again, this has the opposite sense of the separation we observe. A correlation between $T_{dust}$ and $\beta$ might
produce some of the signal we see, but such correlations are notoriously difficult to verify \citep{SHETTY11}. The simplest explanation 
of Figure \ref{fig:tdust} is that an intense ISRF tends to suppress CO emission at a fixed dust abundance, presumably due to enhanced 
dissociation of CO. Enhanced dissociation of CO under a strong ISRF then would lead to reduced area of CO photosphere, likely causing more beam dilution and decreased integrated CO intensity. 
This agrees with the results of \citet{ISRAEL97}, who use dust emission as a tracer of $\rm H_2$ column density to calculate $X_{\rm CO}$,
and find that $X_{\rm CO}$ increases as a function of the strength of interstellar radiation field. 
On the other hand, this disagrees with studies focused on GMC
properties measured from CO emission (\citealt{HUGHES10}, \citealt{PINEDA09}), which find no apparent dependence of the properties of CO clumps or $X_{\rm CO}$ on the strength of radiation field.

\subsection{Median $I_{\rm CO} / A_V$ across different environments}
\label{sec:ico_av_compile}

As a summary of our main results, we compile the median $I_{\rm CO} / A_V$ values across different environments at matched
resolution in Table~\ref{tab:ico_av}. In the MAGMA field, the median $I_{\rm CO} / A_V$ is $\approx 1.9$~K~km~s$^{-1}$~(mag)$^{-1}$.
We find slightly lower values in the APEX field and the SEST field in the SMC, with the median $I_{\rm CO} / A_V$ varies from 1.1 to 1.5~K~km~s$^{-1}$~(mag)$^{-1}$
depending on which transition of CO line to use in the calculation. The median $I_{\rm CO} / A_V$ value for local clouds at matched resolution is $\approx 1.3$~K~km~s$^{-1}$~(mag)$^{-1}$.
The lower $I_{\rm CO} / A_V$ in the integrated measurements of local clouds partly arises from overestimation of $A_V$ in these clouds in the {\em Planck} data (see Section~\ref{sec:int_cloudsdata}).
Also, we note that calculating a median value is sensitive to the area in which  the calculation is made and that cloud boundaries are not known with precision.
If our adopted cloud areas (Table~\ref{tab:lores}) are too large then they may bias the local cloud measurements somewhat low.
As discussed in Section~\ref{sec:mw}, the median $I_{\rm CO} / A_V$ in Milky Way high latitude lines of sight is significantly lower than  
the LMC and SMC, but a correction for contamination by an extended H{\sc i} disk (see Appendix~\ref{sec:appendix_hi_corr} for details) bring these values into closer agreement. 

For comparison, we also recast Galactic CO-to-$\rm H_2$ conversion factor \citep[$X_{\rm CO} = 2 \times 10^{20}$~\xcounits,][]{BOLATTO13} into these units. One can calculate the 
corresponding $I_{\rm CO} / A_V$ for a standard $\xco$, $\approx 4.7$~K~km~s$^{-1}$~(mag)$^{-1}$, by adopting a Galactic dust-to-gas ratio \citep[$N(\rm H) = 5.8 \times 10^{21}~E(B-V)$,][]{BOHLIN78} and $R_V = 3.1$. 
Unlike the rest of the median $I_{\rm CO} / A_V$ values, the recast of $X_{\rm CO}$ only considers molecular gas in the denominator and therefore represents an upper limit to the number that would 
be measured for a real could.

\begin{table}
\begin{center}
\caption{Median $I_{\rm CO}/A_V$ values across environments}
\label{tab:ico_av}
\begin{tabular}{@{}llc@{}}
\hline\hline
Galaxy & Region & $<I_{\rm CO} / A_V>^{a}$  \\
& & (\Kkmpers~mag$^{-1}$) \\
\hline
LMC & MAGMA field & 1.9 \\
& Hot & 1.5 \\
& Cool &  2.2\\
Milky Way & High latitude, without H{\sc i} correction & 0.5\\
& High latitude, with H{\sc i} correction & 0.8 \\
& Local clouds from \em{Planck} & 1.3 \\
& Standard \xco & 4.7\\
SMC & APEX field (CO $J = 2 \to 1$) & 1.1 \\
& SEST field (CO $J = 2 \to 1$) & 1.5\\
& SEST field (CO $J = 1 \to 0$) & 1.4\\
\hline\hline
\end{tabular}
\end{center}
$^{a}$ {We caution that these values are very sensitive to the choice of pixels for computing the values, baseline correction for CO emission in the case of the LMC, and uncertainties associated with deriving $A_V$ from IR emission.}
\end{table}

\section{Discussion}
\label{sec:discussion}
 
\begin{figure}
\includegraphics[width=0.99\linewidth]{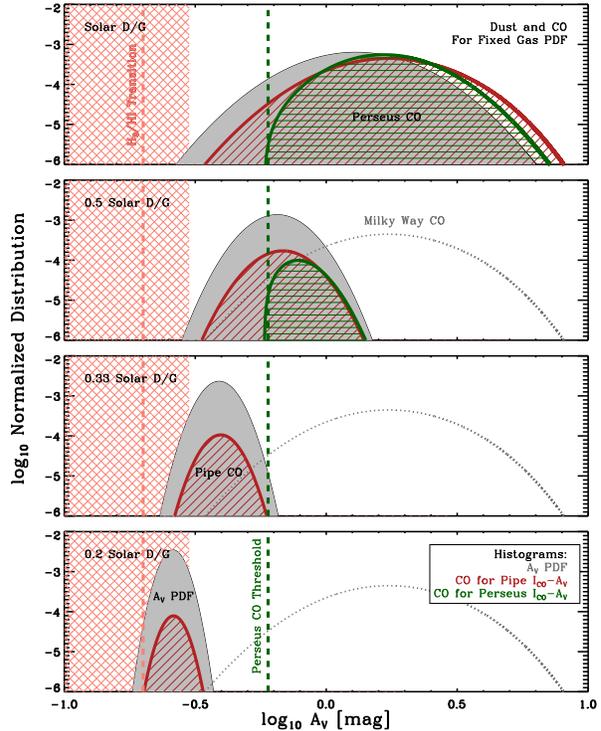}
\caption{\label{fig:sketchpdf} Illustration of our PDF-based calculation of $\xco$. The grey filled histogram shows the same gas column density PDF at a range of
dust-to-gas ratios. The hatched histograms indicate the implied distributions of CO intensity for each of these $A_V$ distributions 
given the Pipe (red diagonal pattern) and Perseus (green horizontal pattern) $I_{\rm CO}$-$A_V$ relations. 
The light red cross hatched region at $A_V \le 0.3$ shows where the transition between H$_2$ and {\sc Hi} will be important. The qualitative difference between Perseus and the Pipe is driven
by the threshold in the Perseus $I_{\rm CO}$-$A_V$ relation, which we illustrate with a green dashed line.
(A colour version of this figure is available in the online journal.)}
\end{figure}

Our comparison of the Magellanic Clouds to the Milky Way argues that on $\sim 10$~pc scales CO intensity 
tracks dust column density, expressed as $A_V$. A logical corollary, though one we cannot prove with existing data,
is that perhaps $I_{\rm CO}$ depends on $A_V$ within clouds in an approximately universal way, at least when averaged over
a sizable population of clouds. If this is the case,
then the integrated CO emission from clouds, and by extension the CO-to-H$_2$ conversion factor, may be thought of as a problem with
several separable parts. First, the distribution of $A_V$ within a cloud will depend on the distribution of gas surface densities
combined with the dust to gas ratio. Second, the CO emission from the cloud will depend on the $A_V$ distribution.
In the case where we are interested in the CO-to-H$_2$ conversion factor, the dust-to-gas ratio and surface density PDF will
also determine what part of the cloud is {\sc Hi} and what part of the cloud is H$_2$. In this simplified view, the CO emission
emerging from a whole cloud can be expressed as

\begin{equation}
\label{eq:ico_cartoon}
I_{\rm CO} = \int_{\rm \Sigma_H} f^{CO,AV} \left( DGR \times PDF\left( \Sigma_H \right)  \right)~,
\end{equation}

\noindent where $\Sigma_H$ is the column density of hydrogen gas; $PDF \left(\Sigma_H \right)$ is the distribution function for gas column densities in the cloud;
$DGR$ is the dust-to-gas ratio, which translated a value of $\Sigma_H$ into $A_V$; and $f^{CO,AV}$ is the function that translates a value
of $A_V$ into a CO intensity. Here the integral is over all values of $\Sigma_H$ and $PDF$ should formally be the total column density 
per differential column density bin.

If one is interested in the CO-to-H$_2$ conversion factor rather than the total CO intensity, then the 

\begin{equation}
\label{eq:xco_cartoon}
\xco \propto \frac{I_{\rm CO}}{N_{\rm H2}} \approx \frac{\int_{\rm \Sigma_H} f^{CO,AV} \left( DGR, \Sigma_H \right)}{\int_{\rm \Sigma_H} f^{\rm H2} \left( \Sigma_H, DGR \right)  PDF\left( \Sigma_H \right)}~.
\end{equation}

\noindent Here the top integral is as above. The bottom integral sums the total H$_2$ mass in the cloud and includes the function $f^{\rm H2} \left(\Sigma_H, DGR \right)$, which indicates
whether at the specified $DGR$ gas of column density $\Sigma_H$ is atomic or molecular.

This very simplified view breaks the CO-to-H$_2$ conversion factor into four parts: the distribution of gas column densities in a cloud, the dust-to-gas ratio, 
the relationship between CO  and dust shielding, and the dependence of the H$_2$/{\sc Hi} transition on $A_V$. The benefit of stating the problem in this simplified way is that each of these
topics has been studied independently (see the Introduction). The last few years have seen substantial work characterizing the column density distribution of local clouds. 
Theoretical work has also established a lognormal distribution as the baseline expectation for a turbulent cloud. Substantial recent theoretical work has also gone in to
understanding the {\sc Hi}-H$_2$ transition. Finally, both theoretical and observational efforts  (including Section \ref{sec:results}) have been made to understand the dependence of $I_{\rm CO}$ 
and $A_V$. That is, this approach breaks the topic of CO emission from clouds in galaxies into a separable problem whose parts may be more tractable than the
topic considered as a whole.
    
\subsection{Implications for the Metallicity Dependence of \xco }
\label{sec:xco}

\begin{figure*}
\includegraphics[width=0.99\linewidth]{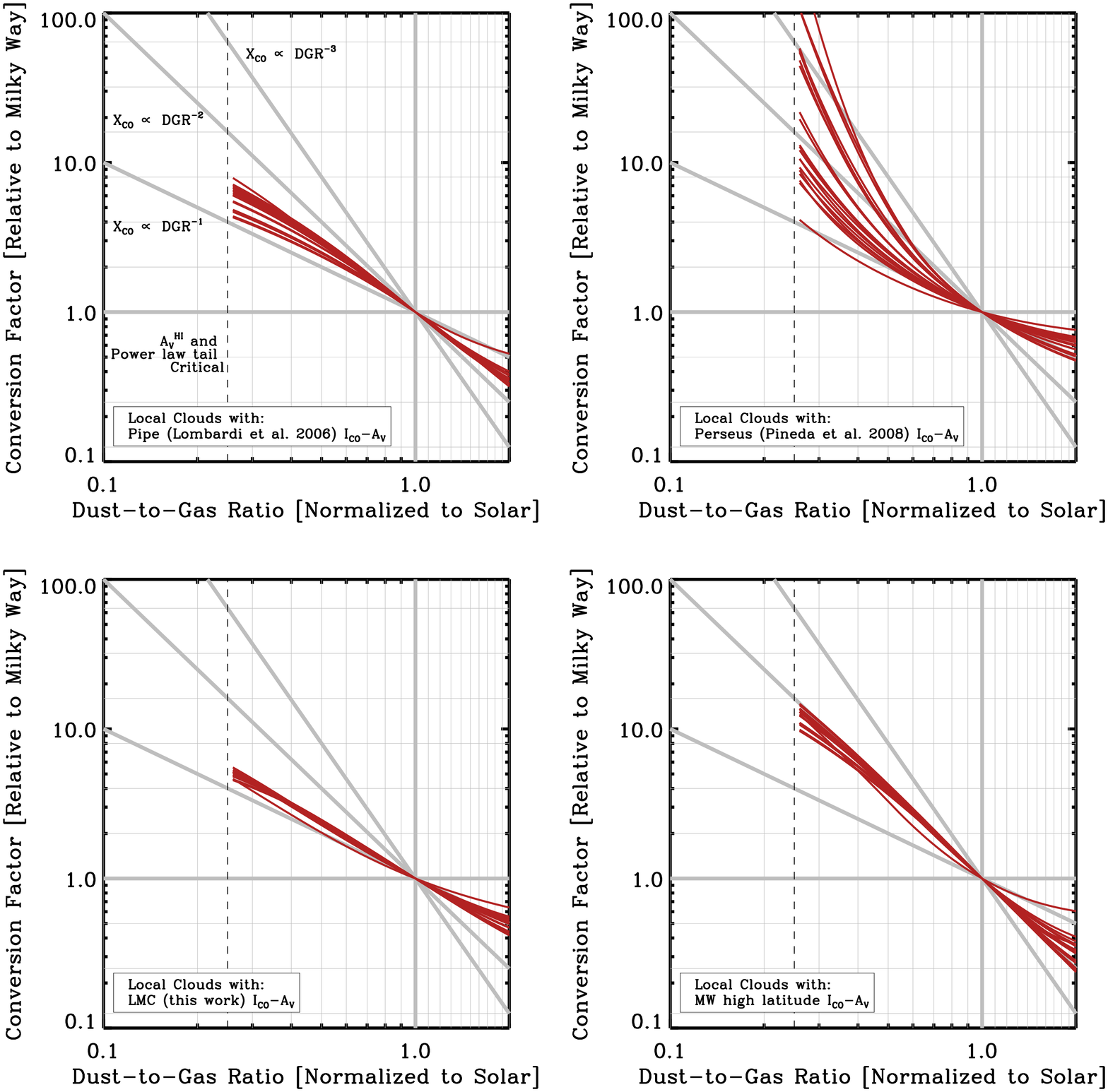}
\caption{\label{fig:xco} $X_{\rm CO}$ as a function of dust-to-gas ratio (and thus metallicity), estimated from the $A_V$ probability distribution function (PDF) observed
for Milky Way clouds \citep{KAIN09} and various $I_{\rm CO}$-$A_V$ relations (Pipe ({\em left top}) : \citealt{LOMBARDI06}, Perseus ({\em right top}) : \citealt{PINEDA08}, LMC ({\em left bottom}), MW high latitude ({\em right bottom}) : polynomial fits to their average $I_{\rm CO}$-$A_V$ relations).
We compute $X_{\rm CO}$ as a function of dust-to-gas ratio following the illustration in Figure~\ref{fig:sketchpdf} (see text for details). 
Each panel shows results for a different functional form of $I_{\rm CO} (A_V)$, while the different red curves in each panel represent the different PDFs from \citet{KAIN09}.
The dashed vertical line denotes a dust-to-gas ratio below which the assumed $A_V$ threshold for the H{\sc i} envelope ($A_{V}^{HI}$) and 
the power law tail in the $A_V$ PDF become critical in the calculation of CO emission.
Thick light grey lines indicate $X_{\rm CO}$ as a function of dust-to-gas ratio with different power law slopes ($X_{\rm CO} \propto$~DGR$^{-\alpha}$ with $\alpha = 1$, 2 and 3).
Just below solar metallicity, $X_{\rm CO}$ is inversely correlated with metallicity with
tractable scatter ($\alpha$ is between 1 and 2). However, by $\sim 1/3$ solar metallicity the input PDF, $I_{\rm CO} (A_V)$ and assumed H{\sc i} layer
dominate the calculation that $X_{\rm CO}$ scatters by two orders of magnitude. Fundamentally this reflects the instability of
using only a very small fraction of the cloud area to trace its total mass and the calculation undermines the utility of CO
to study analogs of local clouds at even moderately sub solar metallicities.
(A colour version of this figure is available in the online journal.)}
\end{figure*}

This sketch can be used to make an empirically-driven prediction for the metallicity dependence of the CO-to-H$_2$ conversion factor, \xco .
If we consider the column density distribution of gas to be universal, then the distribution of $A_V$ is the dust-to-gas ratio times this column density distribution.
This $A_V$ distribution predicts the emergent CO emission from the cloud. The conversion factor depends on the ratio of this CO emission to the sum of the
distribution of gas over the part of the cloud that is molecular. Then by varying the dust-to-gas ratio and repeating the calculation we can derive how
\xco\ changes as a function of $DGR$ in this simple cartoon.

Schematically, Figure \ref{fig:sketchpdf} shows this approach. In this cartoon, gas obeys a universal PDF (here a lognormal), which is scaled by a dust to gas
ratio to yield a PDF of $A_V$ values for each cloud. That PDF appears as a grey normalized histogram. Applying a $A_V$-based prediction for $I_{\rm CO}$, one
arrives at a prediction for the CO intensity. That appears in red and green here for two such functions: the Pipe and the Perseus relation.  This emission would be
summed to get the integrated emission from the cloud. Finally, some part of 
the cloud is atomic (shown by the light red region) and that is not book-kept  in the sum of the molecular mass.

To carry out the calculation quantitatively, we use the parameterization of $A_V$ PDFs for local clouds at solar metallicity listed in Table~1 of \citet{KAIN09}.
In this analysis, we only consider the log-normal part of the $A_V$ PDFs ({\rm i.e.}, ignoring the power-law tail). We also ignore $A_V \le 1$~mag in the 
solar-metallicity clouds because the uncertainty in the extinction mapping used to derive the cloud PDFs becomes substantial below $A_V \sim 1$~mag
(though note that the cartoon in Figure \ref{fig:sketchpdf} does show lower $A_V$ in the top panel; this is just illustrative). These lognormal distributions
clipped at $A_V > 1$~mag are our baseline gas distributions. That is, we consider the gas column density PDF to be:

\begin{equation}
\label{eq:test}
PDF \left( \Sigma_{H,i} (Z_\odot) \right) = \frac{PDF \left(A_{V,i}(Z_\odot) \right)}{DGR(Z_\odot)}
\end{equation}

\noindent where the subscript $i$ refers to one of the \citet{KAIN09} $A_V$ PDFs and $DGR (Z_\odot )$ is the solar metallicity dust-to-gas ratio. Because we
make a relative calculation of \xco , the numerical value of $DGR (Z_\odot )$ will cancel out of our results.

Without specifying what, precisely, $DGR$ is for the Milky Way, we can scale these Milky Way PDFs to those we would expect for otherwise identical clouds
at some fraction of solar metallicity by simply dividing values of $A_V$ by the relative dust-to-gas ratio. That is, by dividing all $A_V$ values by $2$, we 
can shift the PDF to represent an otherwise identical cloud at half solar $DGR$. That is, we hold $PDF \left( \Sigma_{H,i} \right)$ fixed across metallicity and derive
$A_{V,i} \left( Z \right)$ for some sub-solar metallicity, $Z$, via:

\begin{equation}
\label{eq:test2}
PDF \left( A_{V,i}(Z) \right) =  \frac{DGR(Z)}{DGR(Z_\odot)} PDF \left( A_{V,i}(Z_\odot) \right) \approx \frac{Z}{Z_\odot} PDF \left( A_{V,i}(Z_\odot) \right)~.
\end{equation}

\noindent In the last step, we take $Z$ and $DGR$ to vary linearly with one another, but a more complicated dependence \citep[e.g., see][]{REMYRUYER14} 
could easily be introduced into the formulae. Also note that at $Z < Z_\odot$, we will include $A_V < 1$~mag in our calculation; the uncertainty surrounding low
$A_V$ is in the determination of the Milky Way cloud PDFs, not in their inclusion in the calculation.

Next, we translate each $A_V$ PDF into a predicted CO intensity. To do, we input the $A_V$ distribution to one of the 
$I_{\rm CO}$-$A_V$ relationships discussed in the first part of the paper. In this calculation, we consider four relations: 
the Pipe, Perseus, the LMC, and the high latitude Milky Way. For the former two we use the best-fit functions in Table~\ref{tab:highres}. 
For the latter two we fit polynomial functions to the average $I_{\rm CO}$-$A_V$ relations for the LMC and Milky Way high latitude lines of sight 
(red circles and black triangles with error bars in Figure~\ref{fig:mw_avco}, respectively).

Thus we have a set of realistic PDFs, denoted by subscript $i$, and four potential $I_{\rm CO}$-$A_V$ relations, which we denote with the subscript
$j$. We calculate a plausible set of CO intensities emergent from each cloud plus relation $\left( i , j \right)$ for each of a range of metallicities, $Z$:

\begin{equation}
\label{eq:test4}
PDF \left( I_{{\rm CO},i,j}(Z) \right) = f_{j}^{CO,AV}(A_{V,i}(Z))~,
\end{equation}

\noindent where $PDF \left( I_{{\rm CO},i,j}(Z) \right)$ is a PDF of CO intensity at metallicity $Z$ for $A_V$ PDF $i$ and assumed $I_{\rm CO}$-$A_V$ relation $j$.

To estimate \xco , we need to compare the emergent CO intensity to the amount of molecular hydrogen gas, $\Sigma_{{\rm H_2},i}(Z)$. This requires one additional
step, which is to differentiate between H$_2$ and {\sc Hi} in the PDF. \citet{KRUMHOLZ09} and \citet{MCKEE10} argue that the layer of H{\sc i} in a molecular cloud
can be described as a shielding layer of nearly fixed extinction and observations of Perseus support this \citep{LEE12} \citep[see also][]{WOLFIRE10,STERNBERG14}. 
In detail, however, the depth of this layer may vary with metallicity, the radiation field, and volume density \citep[e.g., see][]{WOLFIRE10}. Here we will adopt a simple
approach and adopt a constant $A_V^{\rm HI} \sim 0.2$~mag for each cloud. We take all gas with $A_V$ below this value to arise purely from atomic gas and so
do not book keep it when summing the PDF to obtain an H$_2$ gas mass. A more realistic treatment of this H$_2$-{\sc Hi} transition represents a logical 
extension of this calculation. Here we will only note when this becomes a dominant consideration. 

In this approach

\begin{equation}
\label{eq:test5}
PDF \left( \Sigma_{\rm H2} \right) =
\begin{cases} 
PDF \left( \Sigma_{H} \right) & \mbox{~if~} A_V > A_V^{\rm HI} \\ 
0 & \mbox{~if~} A_V < A_V^{\rm HI} 
\end{cases}
\end{equation}

\noindent where $A_V^{\rm HI}$ is the extinction depth of the {\sc Hi} shielding layer around the cloud.

Combining these equations, we generate $I_{\rm CO}$ and H$_{2}$ distributions as a function of metallicity, cloud structure, and $I_{\rm CO}$-$A_V$ mapping.
Then, the CO-to-H$_2$ conversion factor is simply the total amount of H$_2$ column density divided by the total amount of $I_{\rm CO}$:

\begin{equation}
\label{eq:test6}
\frac{X_{{\rm CO},i,j}(Z)}{X_{{\rm CO},i,j}(Z_\odot)} = \frac{\int PDF \left( \Sigma_{{\rm H_2},i}(Z) \right)}{\int PDF \left( I_{{\rm CO},i,j}(Z) \right)} \frac{\int PDF \left( I_{{\rm CO},i,j}(Z_\odot) \right)}{\int PDF \left( \Sigma_{{\rm H_2},i}(Z_\odot) \right)}
\end{equation}

\noindent By calculating \xco\ relative to the solar metallicity value, the solar metallicity $DGR$ drops out.

Figure \ref{fig:xco} shows the result of the procedure illustrated in Figure \ref{fig:sketchpdf},  
$X_{\rm CO}$ as a function of metallicity (dust-to-gas ratio).
In Figure \ref{fig:xco}, each panel shows the result for the different adopted $I_{\rm CO}$-$A_V$ relation.
In each panel, different curves indicate different $A_V$ PDFs drawn from \citet{KAIN09}.
Gray lines indicate power law dependences of $\xco$ on metallicity for comparison.

This figure illustrates a few points. First, for most adopted PDFs and scalings, between about solar metallicity and $\sim 1/3~Z_\odot$
we expect $\xco \propto Z^{-1}$--$Z^{-2}$, that is, we expect a moderately non-linear scaling in this regime. This is consistent with a number of
theoretical and empirical results summarized in \citet{BOLATTO13}, including results from dust \citep{ISRAEL97,LEROY11} and star formation scaling arguments
\citep{GENZEL12,SCHRUBA12,BLANC13}. It is steeper than most virial mass-based results \citep{WILSON95,ROSOLOWSKY03,LEROY06,BOLATTO08}.
The constraint here can be phrased as follows: if local cloud PDFs are rescaled lower dust-to-gas ratios with no other change in the cloud physics, we might
expect $\xco$ on the scale of whole clouds to scale as $Z^{-1}$ to $Z^{-2}$.

Second, the calculation becomes very sensitive to the H$_2$-{\sc Hi} prescription, the shape of the PDF, and the adopted $I_{\rm CO}$-$A_V$ relation
at low metallicity. Even as high as $1/3~Z_\odot$ these factors create a substantial (factor of $2$) spread among our results. Below this value they dominate 
the results. At some level the spread in our estimates corresponds to a spread in nature, so that this highlights intrinsic scatter or uncertainty in the use of
CO to trace H$_2$ at low metallicities. In this regime, large parts of a cloud may be {\sc Hi}, any threshold for CO emission will become incredibly important,
and any power law tail or extension to high $A_V$ will be preferentially very good at emitting CO. This simple PDF-based approach argues for
intrinsic uncertainties of order a factor of a few when using CO to trace H$_2$ below about $1/3$ solar metallicity.

Finally, though not our focus, the extension of these trends to super-solar metallicity suggests that a factor of $\sim 2$ change in $\xco$ could be achieved
by bringing the Solar Neighborhood clouds to even higher metallicity (and so stronger shielding). At the same time one might expect a number of other changes in the ISM, such as the
emergence of a widespread diffuse molecular phase. But put simply, if all of the molecular gas in the Milky Way were better shielded, as one might expect for 
identical clouds dropped into a more dust-rich system, the conversion factor might be expected to be a factor of $\sim 2$ lower
\citep{PLANCKCOLLABORATIONDARKGAS}. This is not clearly observed \citep[e.g.,][]{DONOVANMEYER13,SANDSTROM13}, but is also not clearly ruled out by observations given that the link between Milky Way and extragalactic
observations remains uncertain at the 10s of percent level.

\subsection{Physics, Key Unknowns, and Complicating Factors in the $I_{\rm CO}$-$A_V$ Relation}
\label{sec:avco_physics}

In calculating $\xco$, the Perseus $I_{\rm CO}$-$A_V$ stands out because it includes a hard $A_V$ threshold
for CO emission. This is particularly stark in Figure \ref{fig:sketchpdf} as below about $1/3$ solar metallicity
almost none of the PDF exceeds the Perseus threshold, suggesting an almost totally CO-dark cloud.
Such a threshold is not visible in our Magellanic Cloud measurements, nor is the saturation
in CO intensities found at high $A_V$ in both the Pipe and Perseus. This is more likely a reflection of our
coarse resolution than the absence of these physical features in the Magellanic Clouds.

In the Appendix, we demonstrate the presence of substantial beam dilution in stacked LMC spectra
by comparing line widths --- which stay about constant --- and peak temperatures --- which drop
to unphysical low levels at low $A_V$. This means that we read our observations results as consistent
with a universal sub-resolution $I_{\rm CO}$-$A_V$ relation, but not as proof of such a relation.
Future measurements comparing dust column density to CO emission across diverse environments
will be needed to establish this relation and its variation at high ($\sim 0.1$~pc, matched to molecular
cloud substructure) resolution. Doing so, key questions will be:

\begin{enumerate}
\item What is the form of the $I_{\rm CO}$-$A_V$ relation at low extinction? Is there a threshold or
steepening of the relation at $\approx 0.5$~mag? Does this also appear in highly resolved maps of
the Magellanic Clouds?
\item How does the $I_{\rm CO}$-$A_V$ change with ambient radiation field? In a first study of this sort,
\citet{INDEBETOUW13} found evidence for suppressed CO emission at low $A_V$ in regions illuminated
by the strong radiation field of 30 Doradus.
\item Is the saturated regime important on the scale of integrals over whole clouds?
\item What is the intrinsic scatter in $I_{\rm CO}$ as a given $A_V$? This captures the degree to which this
one dimensional approach represents a reasonable shorthand for the complex geometry of real clouds.
\end{enumerate}

\noindent A useful goal to enable the sort of calculation we describe above would be a library of relations that
capture the realistic spread among this relation in the Milky Way and Magellanic clouds and allows
for an understanding of how the key features such as the threshold, scatter at fixed $A_V$, saturation level, 
and dependence on environment.

Our knowledge of the PDF requires similar refinement. The PDF of Milky Way clouds at low column remains substantially
unknown \citep{LOMBARDI15}, leading to uncertainties in the functional form of the gas distribution. Similarly, direct
knowledge of the PDF of clouds in other galaxies is almost totally lacking. In the coming years both advances will
help our understanding of the physics of CO emission substantially.

Finally, a careful handling of the different phases of the ISM will improve our understanding of the situation. We have
adopted a very direct observational approach in this paper, simply considering all dust and CO in each $\approx 10$~pc
Magellanic Cloud beam. Via comparison with {\sc Hi} it should eventually be possible to model line of sight contamination by dust 
unassociated with the cloud, though this introduces subtleties regarding what dust is relevant for shielding. An improved 
analytic treatment of the {\sc Hi}-H$_2$ breakdown could also help refine the calculation.

\section{Summary}
\label{sec:summary}

We show that at $10$~pc resolution the relationship between dust column expressed as visual extinction and CO intensity
appears similar in the low metallicity Magellanic Clouds and the Milky Way. This agreement across a range of metallicity 
supports the theoretically motivated view of shielding by dust as the dominant factor in determining the distribution of bright CO emission. To show this,
we use surveys of CO emission from the Large \citep{WONG11} and Small (Rubio et al., in preparation) Magellanic Cloud.
We combine these with estimates of $A_V$ based on {\em Herschel} infrared maps from the 
HERITAGE survey \citep{MEIXNER10}. We compare the Magellanic Cloud measurements to highly resolved Milky Way observations
for two clouds, matched resolution measurements for local molecular clouds, and high latitude CO and dust emission as seen by {\em Planck.}

Our measurements are consistent with an approximately universal relationship between CO intensity and dust extinction within molecular
clouds, though with only $\approx 10$~pc resolution we do not conclusively demonstrate such a relationship. Even for an approximately
universal relation, we still expect such a relationship to vary at second order due to changing geometry and environment. We show suggestive
evidence for such a variation in the Large Magellanic Cloud, where lines of sight with cooler dust temperatures show brighter CO emission
at fixed $A_V$. This could indicate that the weaker radiation field in these regions lowers the density of dissociating photons, allowing
CO to emerge at fainter $A_V$.

We discuss the implications of a nearly universal $I_{\rm CO}$-$A_V$ relationship and suggest a simple, separable model for thinking about
integrated CO emission from molecular clouds. In this picture, the PDF of column densities, the dust-to-gas ratio, the $I_{\rm CO}$-$A_V$ relation,
and the H$_2$-{\sc Hi} boundary combine to determine the properties of a cloud but can be treated as separate problems. A number of
studies have already considered parts of this problem as separable. Here we explore the implications for the CO-to-H$_2$ conversion
factor of a varying dust-to-gas ratio and fixed $I_{\rm CO}$-$A_V$ relation. Taking the PDF of local molecular clouds, we calculate the
corresponding $A_V$ distribution for a range of dust-to-gas ratios and then predict the CO emission for each case. 
The result is a prediction for the variation of CO-to-H$_2$ conversion factor.

Our empirically-motivated model predicts $\xco \propto Z^{-1}$ to $Z^{-2}$ in above about $1/3~Z_\odot$, in rough agreement with
a variety of previous observational and theoretical studies. Our calculation also highlights the tenuous nature of CO as a tracer of
molecular mass at metallicities even as high as $1/5 Z_\odot$. At these metallicities, both the details of the H$_2$-{\sc Hi} transition and 
the shape of the high end of the column density PDF will be extremely important to \xco . For a range of reasonable assumptions our
calculations yield \xco\ that can scatter by as much as an order of magnitude at these metallicities. Future work will be useful to establish 
the functional form and variation of the $I_{\rm CO}$-$A_V$ relation and PDF within clouds, including those at low metallicity and in other Local Group galaxies.

\section*{Acknowledgements}
We thank the anonymous referee for a helpful report. A.D.B. wishes to acknowledge partial support from grants NSF-AST 0955836 and 1412419.
M.R. wishes to acknowledge support from CONICYT(CHILE) through FONDECYTgrant No1140839 and partial support through project BASAL
PFB-06. The National Radio Astronomy Observatory is a facility of the National Science Foundation 
operated under cooperative agreement by Associated Universities, Inc..

\appendix
\section{Effect of Baseline Subtraction, A Thick Galactic H{\sc i} Disk, and 30-Doradus on the $I_{\rm CO}$-$A_V$ relations}
\label{sec:appendix}

\begin{figure*}
\includegraphics[width=0.66\columnwidth]{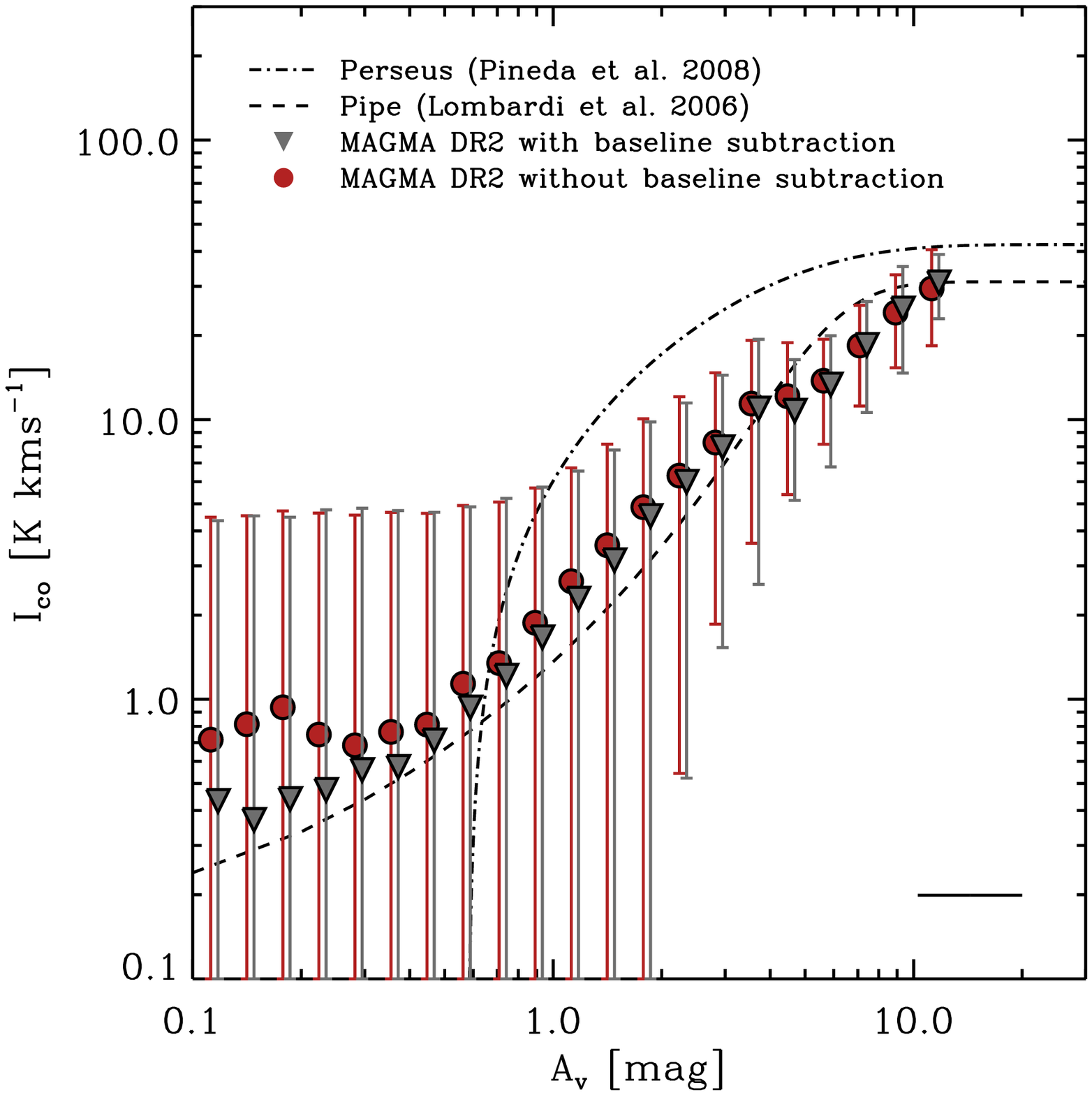}
\includegraphics[width=0.66\columnwidth]{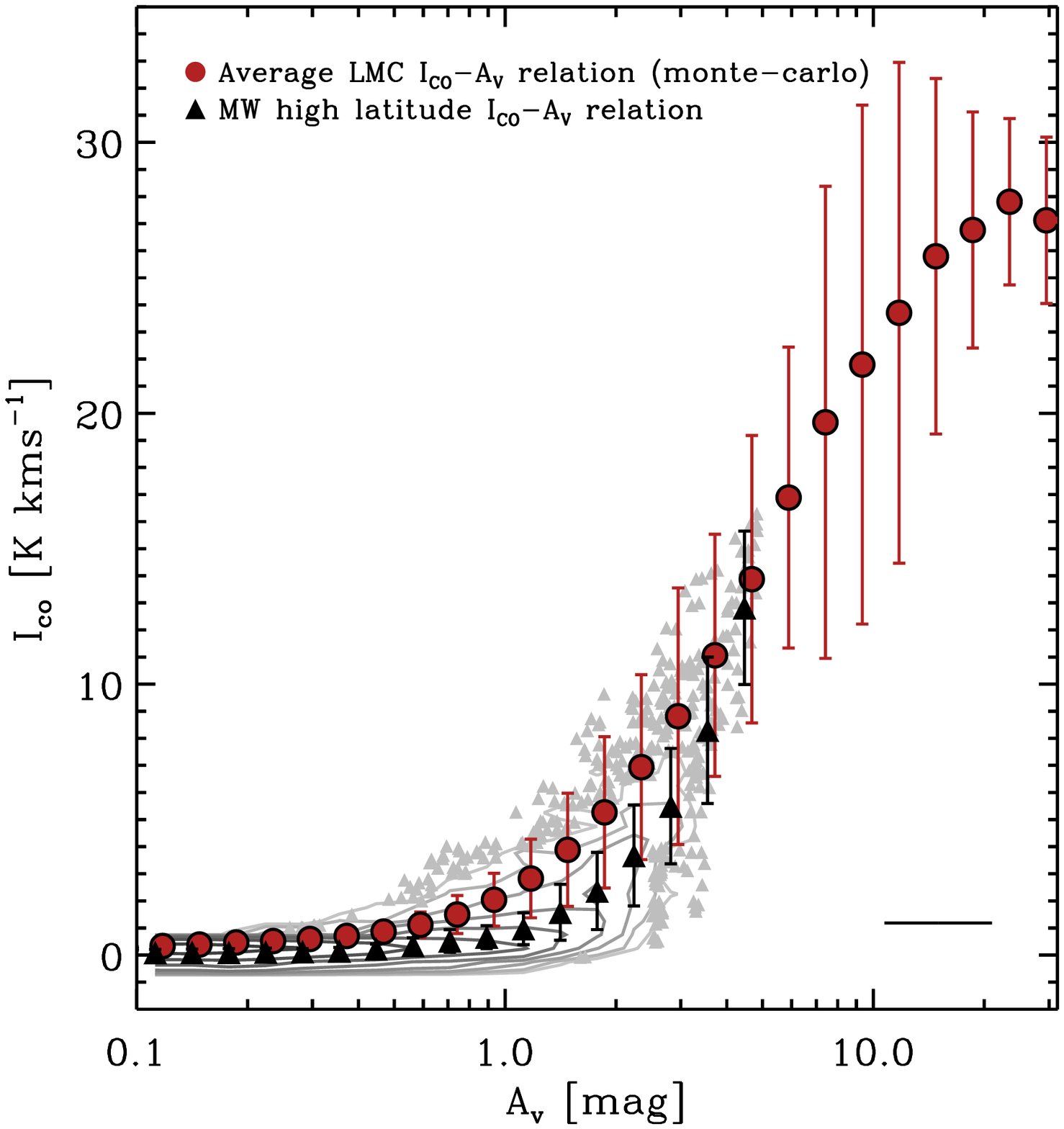}
\includegraphics[width=0.66\columnwidth]{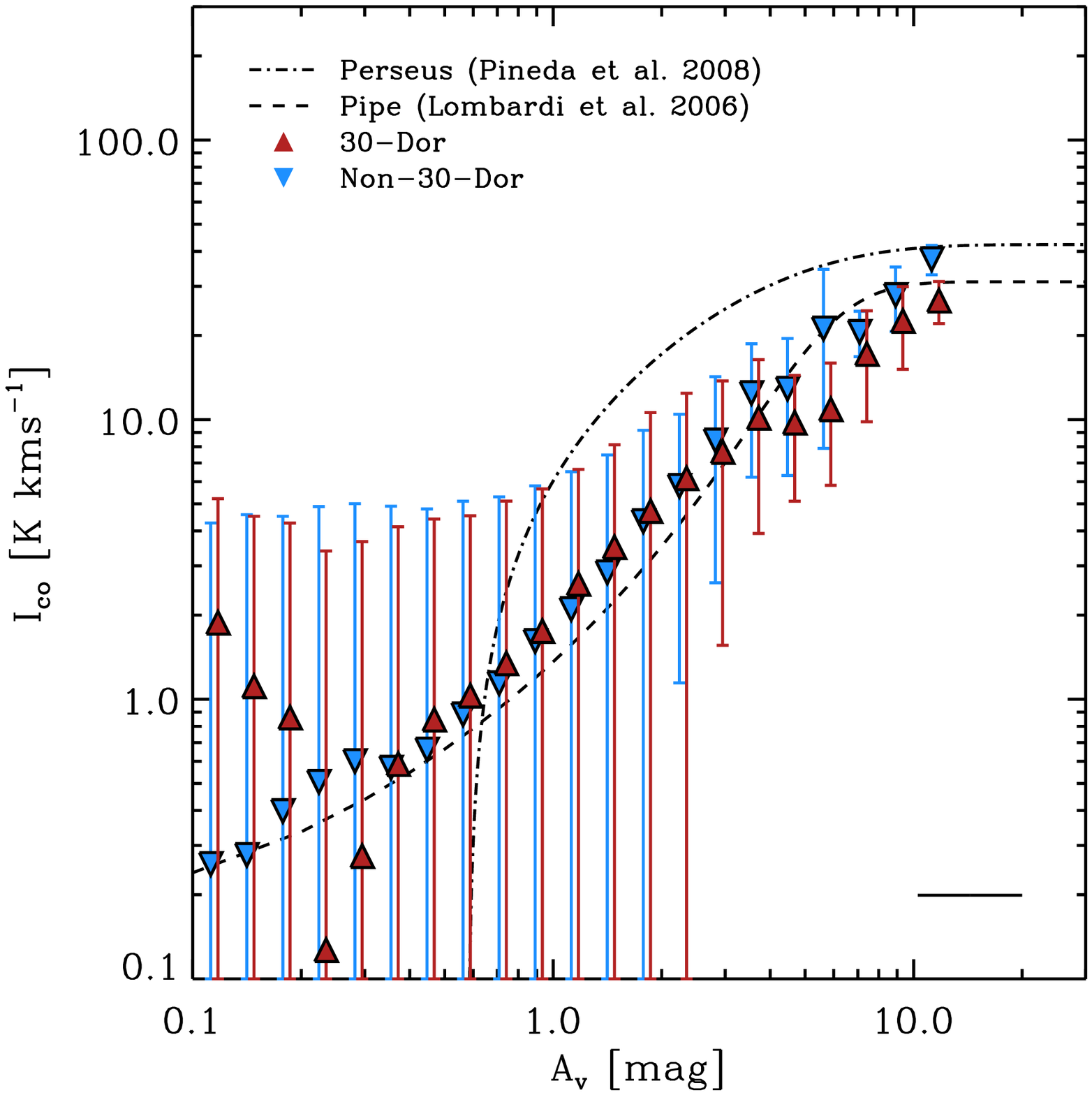}
\caption{\label{fig:appendix} ({\em left}): Effect of baseline subtraction, before (red circles) and after (grey downward triangles; shifted slightly toward right in $x$-axis for easier comparison) our correction. 
Additional baseline subtraction effectively removes the fixed level of CO emission at low $A_V$ while leaving the relation unaffected at high $A_V$. 
({\em middle}): $I_{\rm CO}$-$A_V$ relation in the high latitude Milky Way lines of sight after subtracting the contribution of 400 pc thick H{\sc i} disk to $A_V$ (black triangles). 
The median $I_{\rm CO}$ at a given $A_V$ in the high latitude Milky Way lines of sight is now closer to the average LMC value (red circles), compared to Figure~\ref{fig:mw_avco}. 
({\em right}): $I_{\rm CO}$-$A_V$ relation in the LMC near 30 Doradus (red upward triangles) and outside 30 Doradus (blue downward triangles). 
Although there are differences at low $A_V$, there
is no obvious strong effect associated with being near 30 Doradus above our completeness limit at $A_V \approx 0.8$~mag.
(A colour version of this figure is available in the online journal.)}
\end{figure*}

\subsection{Baseline Subtraction}
\label{sec:appendix_baseline}

In MAGMA DR2, a single linear baseline with magnitude of order a few mK has been subtracted from all spectra. In this paper, we 
carry out an additional additional zeroth order baseline subtraction from the MAGMA data cube pixel-by-pixel. Our local baseline has mean magnitude $\approx 1.2$~mK with $39$~mK 
rms scatter from position to position. Because we integrate the data cube over the whole velocity axis (266 channels with a channel width of $\sim 0.5$ km~s$^{-1}$), this baseline will change the local 
integrated CO intensity by $\approx 0.2 \pm 5$ \Kkmpers. In the left panel of Figure~\ref{fig:appendix}, we show the affect of our additional additional baseline subtraction on the 
$I_{\rm CO}$-$A_V$ relation in the LMC. The additional baseline correction effectively removes the CO emission at low $A_V$, which we thus interpret as likely artifacts.
Overall, though, the relation does not change much and the baseline correction is almost irrelevant at high $A_V$.

\subsection{The Contribution of the Milky Way's Thick H{\sc i} Disk to $A_V$ at High Galactic Latitudes}
\label{sec:appendix_hi_corr}

We argued that the $I_{\rm CO} / A_V$ for the high latitude Milky Way is likely biased low because of contamination by dust associated with a long path length through the 
Milky Way H{\sc i} disk. To estimate the magnitude of this effect, we compute a simple correction factor by adopting a typical hydrogen nuclei number density 
$n \approx 0.1~{\rm  cm}^{-3}$ for the WNM component with a scale height of 400 pc at the location of the Sun (Table~1 in \citealt{Kalberla03}). Assuming a typical
Galactic dust-to-gas ratio, the $A_V$ from this component would be 

\begin{equation}
A_V = 0.1\rm~cm^{-3}\frac{400~{\rm pc} \times 3.09 \times 10^{18}  \rm~cm~{\rm pc}^{-1}}{sin(b)\times1.87\times10^{21}~{\rm cm}^{-2}~{\rm mag}^{-1}},
\end{equation}

\noindent where we convert column density to $A_V$ adopting a Galactic dust-to-gas ratio \citep[$N({\rm H}) = 5.8 \times 10^{21}~E(B-V)$,][]{BOHLIN78} and $R_V = 3.1$, as in Section~\ref{sec:ico_av_compile}.

In the middle panel of Figure~\ref{fig:appendix}, we show $I_{\rm CO}$-$A_V$ relation in the high Galactic latitude lines of sight in the Milky Way after correcting
for the 400 pc thick Milky Way H{\sc i} disk contribution to $A_V$. Compared to Figure~\ref{fig:mw_avco}, the average trend of Milky Way high latitude lines of sight in Figure~\ref{fig:appendix}
is much more closer to the average LMC $I_{\rm CO}$-$A_V$ relation.

\subsection{30-Dor versus Non-30-Dor $I_{\rm CO}$-$A_V$ Relations}
\label{sec:appendix_30dor}

In Section~\ref{sec:tdust} we used the strength of interstellar radiation field traced by $T_{dust}$ as a second parameter to test its effect on $I_{\rm CO}$-$A_V$ relation in the LMC.
We also considered dividing the Large Magellanic Cloud into two regions, 30-Dor and Non-30-Dor (where 30-Dor region is defined as a rectangular box surrounding the H{\sc ii} region), 
to see if there is any systematic effect of star bursting environment on the $I_{\rm CO}$-$A_V$ relation.
The $I_{\rm CO}$-$A_V$ relations in 30-Dor and Non-30-Dor are shown in the right panel of Figure~\ref{fig:appendix}.
Unlike the case for $T_{dust}$, we do not see any notable differences between the $I_{\rm CO}$-$A_V$ relations in 30-Dor and non-30-Dor,
except at very low $A_V$ lines of sight. We suspect that the weird behaviour of the relation at low $A_V$ for 30-Dor region is likely arising from unstable baseline corrections towards CO faint lines of sight. Considering the large dispersion of $I_{\rm CO}$ at a given $A_V$, we conclude that there is no noticeable difference in the $I_{\rm CO}$-$A_V$ relation between the two regions.

\section{Stacked Spectra for the LMC}

\begin{figure*}
\includegraphics[width=0.99\linewidth]{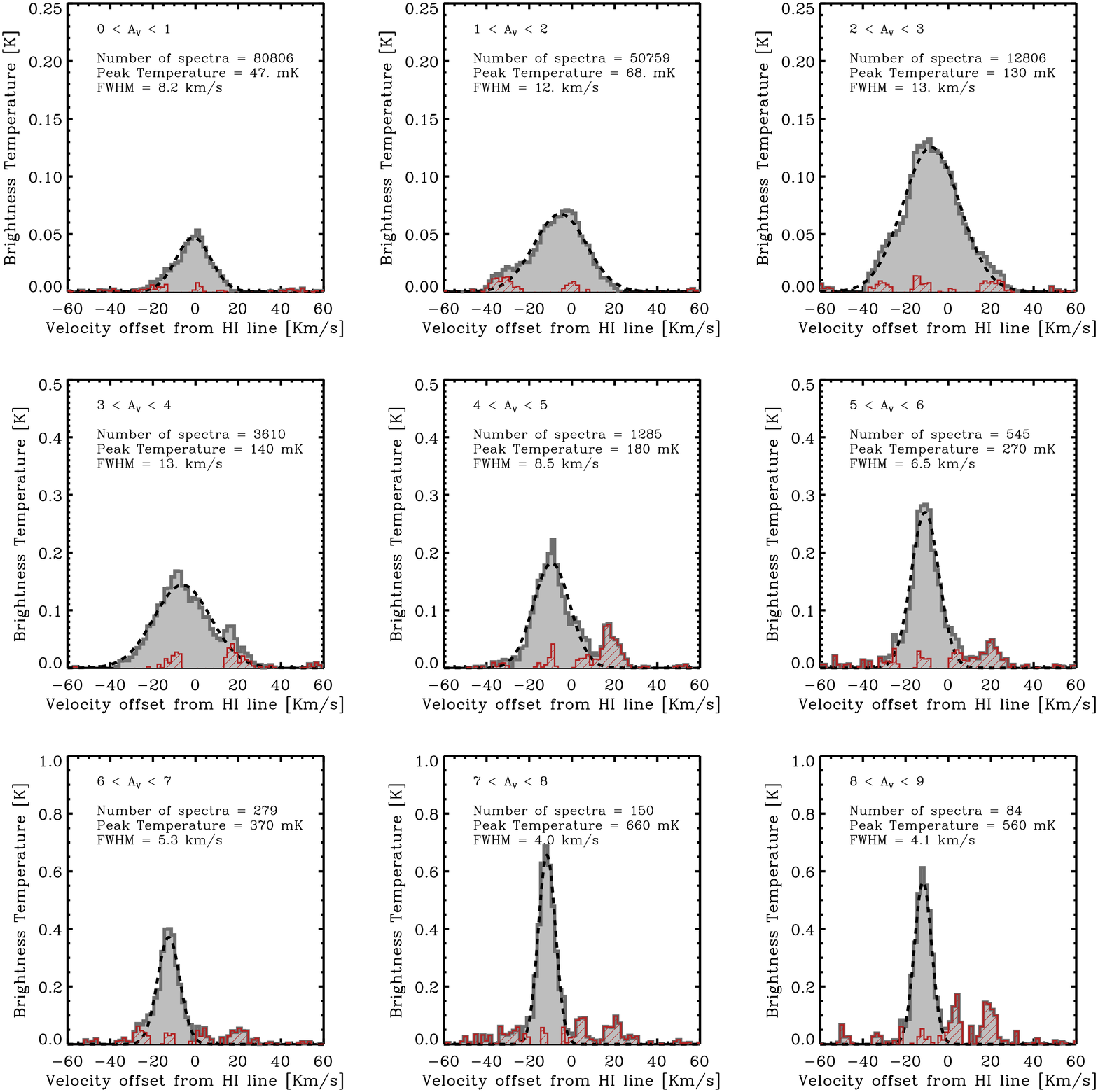}
\caption{\label{fig:spec_tpk_lmc_fit} Stacked CO spectra in bins of $A_V$ in the MAGMA field. Before stacking, we shuffled each spectrum 
using H{\sc i} velocity field from \citet{KIM03} datacube as a template, and fit a single Gaussian to shuffled spectra to estimate the peak brightness temperature ($T_{\rm pk}$) and full width at half maximum (FWHM) of the CO line.
In each panel, grey filled histogram shows stacked spectra, smoothed slightly in velocity axis, and black dashed line shows Gaussian fit to the data. 
Best fit parameters are shown on the left top of each panel, and red diagonally hatched histogram shows the residuals from the fit. 
Significant offset between H{\sc i} and CO velocities of order $\sim$ 20 $\kmpers$ can be identified in the CO 
spectra of high $A_V$ bins, probably due to the fact that H{\sc i} lines are often found with multiple components while 
CO lines usually have single component. In general, we find that the peak brightness temperature increases as a 
function of $A_V$, while there is no clear trend in velocity dispersion. (A colour version of this figure is available in the online journal.)}
\end{figure*}

Our working assumption in the main text of the paper is that similar physics operate at small scales
in molecular clouds in the LMC, SMC, and Milky Way. We interpret the coincidence of LMC, SMC, and Milky
Way data at low resolution in $I_{\rm CO}$-$A_V$ space to support this. A consequence of this conclusion
is that the very low line-integrated intensities seen in the Magellanic Clouds are often a result of dilution by
our 10~pc beam. 

As a basic test of this, we stack the MAGMA spectra in bins of $A_V$ and examine their peak
temperature and line width in each bin. If beam dilution is the dominant physics in setting the observed stacked intensity,
then we expect the peak temperature of the spectrum to vary monotonically with $A_V$ and integrated intensity while the 
line width shows no clear trend. In this case the diminishing intensity simply represents averaging similar spectra with
more and more empty space within the beam. We show that this is the case in Figure~\ref{fig:spec_tpk_lmc_fit}, which plots
CO spectra from LMC in bins of $A_V$. Before averaging, the velocity axis of the CO line has been shuffled 
using H{\sc i} emission \citep{KIM03} as a template \citep[following][]{SCHRUBA11}.

In Figure~\ref{fig:spec_tpk_lmc_fit}, we show the stacked spectrum in each $A_V$ bin as grey filled histograms.
We fit a single Gaussian to each stacked spectrum to derive the peak brightness temperature ($T_{\rm pk}$) 
and full width at half maximum (FWHM) of the CO line. We overplot residuals from the Gaussian fit as red diagonally hatched histograms.
Figure~\ref{fig:spec_tpk_lmc_fit} clearly shows that the peak brightness temperature increases as a function of $A_V$
except in the last $A_V$ bin where a mismatch of CO and H{\sc i} velocity is apparent and leads to lower $T_{\rm pk}$ than the real value.
On the other hand, the fit line width does not have a clear trend. It increases at low $A_V$ and decreases at high $A_V$. 
Therefore, it appears that the peak main beam temperature drives the rise of integrated CO intensity as a function of $A_V$ in the LMC.

The peak values lie far below the expected kinetic temperatures of the molecular gas, so we expect that the dominant
factor in Figure \ref{fig:spec_tpk_lmc_fit} is a changing level of beam dilution. Changing excitation may play some role as
well;  \citet{PINEDA08} found that the CO excitation temperature increases as a function of visual extinction in the
Perseus molecular cloud (See Figure~10 of their paper), ranging from 5~K at low $A_V$ to 20~K at high $A_V$.
Increased heating associated with molecular peaks might also play a role, either due to enhanced star formation 
activity \citep[e.g.,][]{HEIDERMAN10} or efficient photoelectric heating \citep{HUGHES10}. 

Large beam dilution effects has been seen in the Magellanic Clouds as early as \citet{RUBIO93} and \citet{LEQUEUX94}, when it was
invoked to explain the weak CO emission in the SMC. It has also been noted in CO observations of the LMC using SEST telescope \citep{GARAY02,KUTNER97}.
We show that this effect remains strong even at 10~pc resolution. The direct result of this calculation is that the $I_{\rm CO}$-$A_V$ relation
in the Magellanic Clouds must be carefully interpreted. It will not straightforwardly encode features like saturation and an $A_V$ threshold \citep[see also][]{WONG11}.
However, once data are matched in scale, we expect that its similarity to Milky Way results does indicate similar physics operating at sub resolution.

\bibliography{ms}

\begin{thebibliography}{}
\makeatletter
\relax
\def\mn@urlcharsother{\let\do\@makeother \do\$\do\&\do\#\do\^\do\_\do\%\do\~}
\def\mn@doi{\begingroup\mn@urlcharsother \@ifnextchar [ {\mn@doi@}
  {\mn@doi@[]}}
\def\mn@doi@[#1]#2{\def\@tempa{#1}\ifx\@tempa\@empty \href
  {http://dx.doi.org/#2} {doi:#2}\else \href {http://dx.doi.org/#2} {#1}\fi
  \endgroup}
\def\mn@eprint#1#2{\mn@eprint@#1:#2::\@nil}
\def\mn@eprint@arXiv#1{\href {http://arxiv.org/abs/#1} {{\tt arXiv:#1}}}
\def\mn@eprint@dblp#1{\href {http://dblp.uni-trier.de/rec/bibtex/#1.xml}
  {dblp:#1}}
\def\mn@eprint@#1:#2:#3:#4\@nil{\def\@tempa {#1}\def\@tempb {#2}\def\@tempc
  {#3}\ifx \@tempc \@empty \let \@tempc \@tempb \let \@tempb \@tempa \fi \ifx
  \@tempb \@empty \def\@tempb {arXiv}\fi \@ifundefined
  {mn@eprint@\@tempb}{\@tempb:\@tempc}{\expandafter \expandafter \csname
  mn@eprint@\@tempb\endcsname \expandafter{\@tempc}}}

\bibitem[\protect\citeauthoryear{{Aguirre} et~al.,}{{Aguirre}
  et~al.}{2003}]{AGUIRRE03}
{Aguirre} J.~E.,  et~al., 2003, \mn@doi [\apj] {10.1086/377601}, \href
  {http://adsabs.harvard.edu/abs/2003ApJ...596..273A} {596, 273}

\bibitem[\protect\citeauthoryear{{Aniano}, {Draine}, {Gordon}  \&
  {Sandstrom}}{{Aniano} et~al.}{2011}]{ANIANO11}
{Aniano} G.,  {Draine} B.~T.,  {Gordon} K.~D.,   {Sandstrom} K.,  2011, \mn@doi
  [\pasp] {10.1086/662219}, \href
  {http://adsabs.harvard.edu/abs/2011PASP..123.1218A} {123, 1218}

\bibitem[\protect\citeauthoryear{{Bell}, {Roueff}, {Viti}  \&
  {Williams}}{{Bell} et~al.}{2006}]{BELL06}
{Bell} T.~A.,  {Roueff} E.,  {Viti} S.,   {Williams} D.~A.,  2006, \mn@doi
  [\mnras] {10.1111/j.1365-2966.2006.10817.x}, \href
  {http://adsabs.harvard.edu/abs/2006MNRAS.371.1865B} {371, 1865}

\bibitem[\protect\citeauthoryear{{Bernard} et~al.,}{{Bernard}
  et~al.}{2008}]{BERNARD08}
{Bernard} J.-P.,  et~al., 2008, \mn@doi [\aj] {10.1088/0004-6256/136/3/919},
  \href {http://adsabs.harvard.edu/abs/2008AJ....136..919B} {136, 919}

\bibitem[\protect\citeauthoryear{{Blanc} et~al.,}{{Blanc}
  et~al.}{2013}]{BLANC13}
{Blanc} G.~A.,  et~al., 2013, \mn@doi [\apj] {10.1088/0004-637X/764/2/117},
  \href {http://adsabs.harvard.edu/abs/2013ApJ...764..117B} {764, 117}

\bibitem[\protect\citeauthoryear{{Bohlin}, {Savage}  \& {Drake}}{{Bohlin}
  et~al.}{1978}]{BOHLIN78}
{Bohlin} R.~C.,  {Savage} B.~D.,   {Drake} J.~F.,  1978, \mn@doi [\apj]
  {10.1086/156357}, \href {http://adsabs.harvard.edu/abs/1978ApJ...224..132B}
  {224, 132}

\bibitem[\protect\citeauthoryear{{Bolatto}, {Leroy}, {Israel}  \&
  {Jackson}}{{Bolatto} et~al.}{2003}]{BOLATTO03}
{Bolatto} A.~D.,  {Leroy} A.,  {Israel} F.~P.,   {Jackson} J.~M.,  2003,
  \mn@doi [\apj] {10.1086/377230}, \href
  {http://adsabs.harvard.edu/abs/2003ApJ...595..167B} {595, 167}

\bibitem[\protect\citeauthoryear{{Bolatto}, {Leroy}, {Rosolowsky}, {Walter}  \&
  {Blitz}}{{Bolatto} et~al.}{2008}]{BOLATTO08}
{Bolatto} A.~D.,  {Leroy} A.~K.,  {Rosolowsky} E.,  {Walter} F.,   {Blitz} L.,
  2008, \mn@doi [\apj] {10.1086/591513}, \href
  {http://adsabs.harvard.edu/abs/2008ApJ...686..948B} {686, 948}

\bibitem[\protect\citeauthoryear{{Bolatto}, {Wolfire}  \& {Leroy}}{{Bolatto}
  et~al.}{2013}]{BOLATTO13}
{Bolatto} A.~D.,  {Wolfire} M.,   {Leroy} A.~K.,  2013, \mn@doi [\araa]
  {10.1146/annurev-astro-082812-140944}, \href
  {http://adsabs.harvard.edu/abs/2013ARA%26A..51..207B} {51, 207}

\bibitem[\protect\citeauthoryear{{Boulanger}, {Abergel}, {Bernard}, {Burton},
  {Desert}, {Hartmann}, {Lagache}  \& {Puget}}{{Boulanger}
  et~al.}{1996}]{1996A&A...312..256B}
{Boulanger} F.,  {Abergel} A.,  {Bernard} J.-P.,  {Burton} W.~B.,  {Desert}
  F.-X.,  {Hartmann} D.,  {Lagache} G.,   {Puget} J.-L.,  1996, \aap, \href
  {http://adsabs.harvard.edu/abs/1996A%26A...312..256B} {312, 256}

\bibitem[\protect\citeauthoryear{{Dame}, {Hartmann}  \& {Thaddeus}}{{Dame}
  et~al.}{2001}]{DAME01}
{Dame} T.~M.,  {Hartmann} D.,   {Thaddeus} P.,  2001, \mn@doi [\apj]
  {10.1086/318388}, \href {http://adsabs.harvard.edu/abs/2001ApJ...547..792D}
  {547, 792}

\bibitem[\protect\citeauthoryear{{Dobashi}, {Bernard}, {Hughes}, {Paradis},
  {Reach}  \& {Kawamura}}{{Dobashi} et~al.}{2008}]{DOBASHI08}
{Dobashi} K.,  {Bernard} J.-P.,  {Hughes} A.,  {Paradis} D.,  {Reach} W.~T.,
  {Kawamura} A.,  2008, \mn@doi [\aap] {10.1051/0004-6361:20079151}, \href
  {http://adsabs.harvard.edu/abs/2008A%26A...484..205D} {484, 205}

\bibitem[\protect\citeauthoryear{{Donovan Meyer} et~al.,}{{Donovan Meyer}
  et~al.}{2013}]{DONOVANMEYER13}
{Donovan Meyer} J.,  et~al., 2013, \mn@doi [\apj]
  {10.1088/0004-637X/772/2/107}, \href
  {http://adsabs.harvard.edu/abs/2013ApJ...772..107D} {772, 107}

\bibitem[\protect\citeauthoryear{{Downes} \& {Solomon}}{{Downes} \&
  {Solomon}}{1998}]{DOWNES98}
{Downes} D.,  {Solomon} P.~M.,  1998, \mn@doi [\apj] {10.1086/306339}, \href
  {http://adsabs.harvard.edu/abs/1998ApJ...507..615D} {507, 615}

\bibitem[\protect\citeauthoryear{{Draine} \& {Lee}}{{Draine} \&
  {Lee}}{1984}]{DL84}
{Draine} B.~T.,  {Lee} H.~M.,  1984, \mn@doi [\apj] {10.1086/162480}, \href
  {http://adsabs.harvard.edu/abs/1984ApJ...285...89D} {285, 89}

\bibitem[\protect\citeauthoryear{{Draine} et~al.,}{{Draine}
  et~al.}{2007}]{DRAINE07}
{Draine} B.~T.,  et~al., 2007, \mn@doi [\apj] {10.1086/518306}, \href
  {http://adsabs.harvard.edu/abs/2007ApJ...663..866D} {663, 866}

\bibitem[\protect\citeauthoryear{{Dupac} et~al.,}{{Dupac}
  et~al.}{2003}]{DUPAC03}
{Dupac} X.,  et~al., 2003, \mn@doi [\aap] {10.1051/0004-6361:20030575}, \href
  {http://adsabs.harvard.edu/abs/2003A%26A...404L..11D} {404, L11}

\bibitem[\protect\citeauthoryear{{Fukui} et~al.,}{{Fukui}
  et~al.}{1999}]{FUKUI99}
{Fukui} Y.,  et~al., 1999, \pasj, \href
  {http://adsabs.harvard.edu/abs/1999PASJ...51..745F} {51, 745}

\bibitem[\protect\citeauthoryear{{Fukui} et~al.,}{{Fukui}
  et~al.}{2008}]{FUKUI08}
{Fukui} Y.,  et~al., 2008, \mn@doi [\apjs] {10.1086/589833}, \href
  {http://adsabs.harvard.edu/abs/2008ApJS..178...56F} {178, 56}

\bibitem[\protect\citeauthoryear{{Garay}, {Johansson}, {Nyman}, {Booth},
  {Israel}, {Kutner}, {Lequeux}  \& {Rubio}}{{Garay} et~al.}{2002}]{GARAY02}
{Garay} G.,  {Johansson} L.~E.~B.,  {Nyman} L.-{\AA}.,  {Booth} R.~S.,
  {Israel} F.~P.,  {Kutner} M.~L.,  {Lequeux} J.,   {Rubio} M.,  2002, \mn@doi
  [\aap] {10.1051/0004-6361:20020397}, \href
  {http://adsabs.harvard.edu/abs/2002A%26A...389..977G} {389, 977}

\bibitem[\protect\citeauthoryear{{Genzel} et~al.,}{{Genzel}
  et~al.}{2012}]{GENZEL12}
{Genzel} R.,  et~al., 2012, \mn@doi [\apj] {10.1088/0004-637X/746/1/69}, \href
  {http://adsabs.harvard.edu/abs/2012ApJ...746...69G} {746, 69}

\bibitem[\protect\citeauthoryear{{Glover} \& {Mac Low}}{{Glover} \& {Mac
  Low}}{2011}]{GLOVER11}
{Glover} S.~C.~O.,  {Mac Low} M.-M.,  2011, \mn@doi [\mnras]
  {10.1111/j.1365-2966.2010.17907.x}, \href
  {http://adsabs.harvard.edu/abs/2011MNRAS.412..337G} {412, 337}

\bibitem[\protect\citeauthoryear{{Gordon} et~al.,}{{Gordon}
  et~al.}{2010}]{GORDON10}
{Gordon} K.~D.,  et~al., 2010, \mn@doi [\aap] {10.1051/0004-6361/201014541},
  \href {http://adsabs.harvard.edu/abs/2010A%26A...518L..89G} {518, L89}

\bibitem[\protect\citeauthoryear{{Heiderman}, {Evans}, {Allen}, {Huard}  \&
  {Heyer}}{{Heiderman} et~al.}{2010}]{HEIDERMAN10}
{Heiderman} A.,  {Evans} II N.~J.,  {Allen} L.~E.,  {Huard} T.,   {Heyer} M.,
  2010, \mn@doi [\apj] {10.1088/0004-637X/723/2/1019}, \href
  {http://adsabs.harvard.edu/abs/2010ApJ...723.1019H} {723, 1019}

\bibitem[\protect\citeauthoryear{{Hildebrand}}{{Hildebrand}}{1983}]{HILDEBRAND83}
{Hildebrand} R.~H.,  1983, \qjras, \href
  {http://adsabs.harvard.edu/abs/1983QJRAS..24..267H} {24, 267}

\bibitem[\protect\citeauthoryear{{Hughes} et~al.,}{{Hughes}
  et~al.}{2010}]{HUGHES10}
{Hughes} A.,  et~al., 2010, \mn@doi [\mnras]
  {10.1111/j.1365-2966.2010.16829.x}, \href
  {http://adsabs.harvard.edu/abs/2010MNRAS.406.2065H} {406, 2065}

\bibitem[\protect\citeauthoryear{{Indebetouw} et~al.,}{{Indebetouw}
  et~al.}{2013}]{INDEBETOUW13}
{Indebetouw} R.,  et~al., 2013, \mn@doi [\apj] {10.1088/0004-637X/774/1/73},
  \href {http://adsabs.harvard.edu/abs/2013ApJ...774...73I} {774, 73}

\bibitem[\protect\citeauthoryear{{Israel}}{{Israel}}{1997}]{ISRAEL97}
{Israel} F.~P.,  1997, \aap, \href
  {http://adsabs.harvard.edu/abs/1997A%26A...328..471I} {328, 471}

\bibitem[\protect\citeauthoryear{{Kainulainen}, {Beuther}, {Henning}  \&
  {Plume}}{{Kainulainen} et~al.}{2009}]{KAIN09}
{Kainulainen} J.,  {Beuther} H.,  {Henning} T.,   {Plume} R.,  2009, \mn@doi
  [\aap] {10.1051/0004-6361/200913605}, \href
  {http://adsabs.harvard.edu/abs/2009A%26A...508L..35K} {508, L35}

\bibitem[\protect\citeauthoryear{{Kalberla}}{{Kalberla}}{2003}]{Kalberla03}
{Kalberla} P.~M.~W.,  2003, \mn@doi [\apj] {10.1086/374330}, \href
  {http://adsabs.harvard.edu/abs/2003ApJ...588..805K} {588, 805}

\bibitem[\protect\citeauthoryear{{Keller} \& {Wood}}{{Keller} \&
  {Wood}}{2006}]{KELLER06}
{Keller} S.~C.,  {Wood} P.~R.,  2006, \mn@doi [\apj] {10.1086/501115}, \href
  {http://adsabs.harvard.edu/abs/2006ApJ...642..834K} {642, 834}

\bibitem[\protect\citeauthoryear{{Kennicutt} \& {Evans}}{{Kennicutt} \&
  {Evans}}{2012}]{KENNICUTT12}
{Kennicutt} R.~C.,  {Evans} N.~J.,  2012, \mn@doi [\araa]
  {10.1146/annurev-astro-081811-125610}, \href
  {http://adsabs.harvard.edu/abs/2012ARA%26A..50..531K} {50, 531}

\bibitem[\protect\citeauthoryear{{Kim}, {Staveley-Smith}, {Dopita}, {Sault},
  {Freeman}, {Lee}  \& {Chu}}{{Kim} et~al.}{2003}]{KIM03}
{Kim} S.,  {Staveley-Smith} L.,  {Dopita} M.~A.,  {Sault} R.~J.,  {Freeman}
  K.~C.,  {Lee} Y.,   {Chu} Y.-H.,  2003, \mn@doi [\apjs] {10.1086/376980},
  \href {http://adsabs.harvard.edu/abs/2003ApJS..148..473K} {148, 473}

\bibitem[\protect\citeauthoryear{{Krumholz}, {McKee}  \&
  {Tumlinson}}{{Krumholz} et~al.}{2009}]{KRUMHOLZ09}
{Krumholz} M.~R.,  {McKee} C.~F.,   {Tumlinson} J.,  2009, \mn@doi [\apj]
  {10.1088/0004-637X/693/1/216}, \href
  {http://adsabs.harvard.edu/abs/2009ApJ...693..216K} {693, 216}

\bibitem[\protect\citeauthoryear{{Kutner} et~al.,}{{Kutner}
  et~al.}{1997}]{KUTNER97}
{Kutner} M.~L.,  et~al., 1997, \mn@doi [\aaps] {10.1051/aas:1997334}, \href
  {http://adsabs.harvard.edu/abs/1997A%26AS..122..255K} {122, 255}

\bibitem[\protect\citeauthoryear{{Lee} et~al.,}{{Lee} et~al.}{2012}]{LEE12}
{Lee} M.-Y.,  et~al., 2012, \mn@doi [\apj] {10.1088/0004-637X/748/2/75}, \href
  {http://adsabs.harvard.edu/abs/2012ApJ...748...75L} {748, 75}

\bibitem[\protect\citeauthoryear{{Lequeux}, {Le Bourlot}, {Pineau des Forets},
  {Roueff}, {Boulanger}  \& {Rubio}}{{Lequeux} et~al.}{1994}]{LEQUEUX94}
{Lequeux} J.,  {Le Bourlot} J.,  {Pineau des Forets} G.,  {Roueff} E.,
  {Boulanger} F.,   {Rubio} M.,  1994, \aap, \href
  {http://adsabs.harvard.edu/abs/1994A%26A...292..371L} {292, 371}

\bibitem[\protect\citeauthoryear{{Leroy}, {Bolatto}, {Walter}  \&
  {Blitz}}{{Leroy} et~al.}{2006}]{LEROY06}
{Leroy} A.,  {Bolatto} A.,  {Walter} F.,   {Blitz} L.,  2006, \mn@doi [\apj]
  {10.1086/503024}, \href {http://adsabs.harvard.edu/abs/2006ApJ...643..825L}
  {643, 825}

\bibitem[\protect\citeauthoryear{{Leroy}, {Bolatto}, {Stanimirovic}, {Mizuno},
  {Israel}  \& {Bot}}{{Leroy} et~al.}{2007}]{LEROY07}
{Leroy} A.,  {Bolatto} A.,  {Stanimirovic} S.,  {Mizuno} N.,  {Israel} F.,
  {Bot} C.,  2007, \mn@doi [\apj] {10.1086/511150}, \href
  {http://adsabs.harvard.edu/abs/2007ApJ...658.1027L} {658, 1027}

\bibitem[\protect\citeauthoryear{{Leroy} et~al.,}{{Leroy}
  et~al.}{2009}]{LEROY09}
{Leroy} A.~K.,  et~al., 2009, \mn@doi [\apj] {10.1088/0004-637X/702/1/352},
  \href {http://adsabs.harvard.edu/abs/2009ApJ...702..352L} {702, 352}

\bibitem[\protect\citeauthoryear{{Leroy} et~al.,}{{Leroy}
  et~al.}{2011}]{LEROY11}
{Leroy} A.~K.,  et~al., 2011, \mn@doi [\apj] {10.1088/0004-637X/737/1/12},
  \href {http://adsabs.harvard.edu/abs/2011ApJ...737...12L} {737, 12}

\bibitem[\protect\citeauthoryear{{Lombardi}, {Alves}  \& {Lada}}{{Lombardi}
  et~al.}{2006}]{LOMBARDI06}
{Lombardi} M.,  {Alves} J.,   {Lada} C.~J.,  2006, \mn@doi [\aap]
  {10.1051/0004-6361:20042474}, \href
  {http://adsabs.harvard.edu/abs/2006A%26A...454..781L} {454, 781}

\bibitem[\protect\citeauthoryear{{Lombardi}, {Alves}  \& {Lada}}{{Lombardi}
  et~al.}{2015}]{LOMBARDI15}
{Lombardi} M.,  {Alves} J.,   {Lada} C.~J.,  2015, preprint, \href
  {http://adsabs.harvard.edu/abs/2015arXiv150203859L} {} (\mn@eprint {arXiv}
  {1502.03859})

\bibitem[\protect\citeauthoryear{{Maloney} \& {Black}}{{Maloney} \&
  {Black}}{1988}]{MB88}
{Maloney} P.,  {Black} J.~H.,  1988, \mn@doi [\apj] {10.1086/166011}, \href
  {http://adsabs.harvard.edu/abs/1988ApJ...325..389M} {325, 389}

\bibitem[\protect\citeauthoryear{{McKee} \& {Krumholz}}{{McKee} \&
  {Krumholz}}{2010}]{MCKEE10}
{McKee} C.~F.,  {Krumholz} M.~R.,  2010, \mn@doi [\apj]
  {10.1088/0004-637X/709/1/308}, \href
  {http://adsabs.harvard.edu/abs/2010ApJ...709..308M} {709, 308}

\bibitem[\protect\citeauthoryear{{Meixner} et~al.,}{{Meixner}
  et~al.}{2006}]{MEIXNER06}
{Meixner} M.,  et~al., 2006, \mn@doi [\aj] {10.1086/508185}, \href
  {http://adsabs.harvard.edu/abs/2006AJ....132.2268M} {132, 2268}

\bibitem[\protect\citeauthoryear{{Meixner} et~al.,}{{Meixner}
  et~al.}{2010}]{MEIXNER10}
{Meixner} M.,  et~al., 2010, \mn@doi [\aap] {10.1051/0004-6361/201014662},
  \href {http://adsabs.harvard.edu/abs/2010A%26A...518L..71M} {518, L71}

\bibitem[\protect\citeauthoryear{{Meixner} et~al.,}{{Meixner}
  et~al.}{2013}]{MEIXNER13}
{Meixner} M.,  et~al., 2013, \mn@doi [\aj] {10.1088/0004-6256/146/3/62}, \href
  {http://adsabs.harvard.edu/abs/2013AJ....146...62M} {146, 62}

\bibitem[\protect\citeauthoryear{{Narayanan}, {Krumholz}, {Ostriker}  \&
  {Hernquist}}{{Narayanan} et~al.}{2012}]{NARAYANAN12}
{Narayanan} D.,  {Krumholz} M.~R.,  {Ostriker} E.~C.,   {Hernquist} L.,  2012,
  \mn@doi [\mnras] {10.1111/j.1365-2966.2012.20536.x}, \href
  {http://adsabs.harvard.edu/abs/2012MNRAS.421.3127N} {421, 3127}

\bibitem[\protect\citeauthoryear{{Pineda}, {Caselli}  \& {Goodman}}{{Pineda}
  et~al.}{2008}]{PINEDA08}
{Pineda} J.~E.,  {Caselli} P.,   {Goodman} A.~A.,  2008, \mn@doi [\apj]
  {10.1086/586883}, \href {http://adsabs.harvard.edu/abs/2008ApJ...679..481P}
  {679, 481}

\bibitem[\protect\citeauthoryear{{Pineda}, {Ott}, {Klein}, {Wong}, {Muller}  \&
  {Hughes}}{{Pineda} et~al.}{2009}]{PINEDA09}
{Pineda} J.~L.,  {Ott} J.,  {Klein} U.,  {Wong} T.,  {Muller} E.,   {Hughes}
  A.,  2009, \mn@doi [\apj] {10.1088/0004-637X/703/1/736}, \href
  {http://adsabs.harvard.edu/abs/2009ApJ...703..736P} {703, 736}

\bibitem[\protect\citeauthoryear{{Pineda}, {Goldsmith}, {Chapman}, {Snell},
  {Li}, {Cambr{\'e}sy}  \& {Brunt}}{{Pineda} et~al.}{2010}]{PINEDA10}
{Pineda} J.~L.,  {Goldsmith} P.~F.,  {Chapman} N.,  {Snell} R.~L.,  {Li} D.,
  {Cambr{\'e}sy} L.,   {Brunt} C.,  2010, \mn@doi [\apj]
  {10.1088/0004-637X/721/1/686}, \href
  {http://adsabs.harvard.edu/abs/2010ApJ...721..686P} {721, 686}

\bibitem[\protect\citeauthoryear{{Planck Collaboration} et~al.,}{{Planck
  Collaboration} et~al.}{2011a}]{PLANCK11}
{Planck Collaboration} et~al., 2011a, \mn@doi [\aap]
  {10.1051/0004-6361/201116473}, \href
  {http://adsabs.harvard.edu/abs/2011A%26A...536A..17P} {536, A17}

\bibitem[\protect\citeauthoryear{{Planck Collaboration} et~al.,}{{Planck
  Collaboration} et~al.}{2011b}]{PLANCKCOLLABORATIONDARKGAS}
{Planck Collaboration} et~al., 2011b, \mn@doi [\aap]
  {10.1051/0004-6361/201116479}, \href
  {http://adsabs.harvard.edu/abs/2011A%26A...536A..19P} {536, A19}

\bibitem[\protect\citeauthoryear{{Planck Collaboration} et~al.,}{{Planck
  Collaboration} et~al.}{2011c}]{PLANCK11_MC}
{Planck Collaboration} et~al., 2011c, \mn@doi [\aap]
  {10.1051/0004-6361/201116483}, \href
  {http://adsabs.harvard.edu/abs/2011A%26A...536A..25P} {536, A25}

\bibitem[\protect\citeauthoryear{{Planck Collaboration} et~al.,}{{Planck
  Collaboration} et~al.}{2013b}]{PLANCK13_CO}
{Planck Collaboration} et~al., 2013b, preprint, \href
  {http://adsabs.harvard.edu/abs/2013arXiv1303.5073P} {} (\mn@eprint {arXiv}
  {1303.5073})

\bibitem[\protect\citeauthoryear{{Planck Collaboration} et~al.,}{{Planck
  Collaboration} et~al.}{2013a}]{PLANCK13_DUST}
{Planck Collaboration} et~al., 2013a, preprint, \href
  {http://adsabs.harvard.edu/abs/2013arXiv1312.1300P} {} (\mn@eprint {arXiv}
  {1312.1300})

\bibitem[\protect\citeauthoryear{{R{\'e}my-Ruyer} et~al.,}{{R{\'e}my-Ruyer}
  et~al.}{2014}]{REMYRUYER14}
{R{\'e}my-Ruyer} A.,  et~al., 2014, \mn@doi [\aap]
  {10.1051/0004-6361/201322803}, \href
  {http://adsabs.harvard.edu/abs/2014A%26A...563A..31R} {563, A31}

\bibitem[\protect\citeauthoryear{{Rosolowsky}, {Engargiola}, {Plambeck}  \&
  {Blitz}}{{Rosolowsky} et~al.}{2003}]{ROSOLOWSKY03}
{Rosolowsky} E.,  {Engargiola} G.,  {Plambeck} R.,   {Blitz} L.,  2003, \mn@doi
  [\apj] {10.1086/379166}, \href
  {http://adsabs.harvard.edu/abs/2003ApJ...599..258R} {599, 258}

\bibitem[\protect\citeauthoryear{{Rubio}, {Lequeux}  \& {Boulanger}}{{Rubio}
  et~al.}{1993}]{RUBIO93}
{Rubio} M.,  {Lequeux} J.,   {Boulanger} F.,  1993, \aap, \href
  {http://adsabs.harvard.edu/abs/1993A%26A...271....9R} {271, 9}

\bibitem[\protect\citeauthoryear{{Sandstrom} et~al.,}{{Sandstrom}
  et~al.}{2013}]{SANDSTROM13}
{Sandstrom} K.~M.,  et~al., 2013, \mn@doi [\apj] {10.1088/0004-637X/777/1/5},
  \href {http://adsabs.harvard.edu/abs/2013ApJ...777....5S} {777, 5}

\bibitem[\protect\citeauthoryear{{Schlegel}, {Finkbeiner}  \&
  {Davis}}{{Schlegel} et~al.}{1998}]{SFD98}
{Schlegel} D.~J.,  {Finkbeiner} D.~P.,   {Davis} M.,  1998, \mn@doi [\apj]
  {10.1086/305772}, \href {http://adsabs.harvard.edu/abs/1998ApJ...500..525S}
  {500, 525}

\bibitem[\protect\citeauthoryear{{Schnee}, {Ridge}, {Goodman}  \&
  {Li}}{{Schnee} et~al.}{2005}]{2005ApJ...634..442S}
{Schnee} S.~L.,  {Ridge} N.~A.,  {Goodman} A.~A.,   {Li} J.~G.,  2005, \mn@doi
  [\apj] {10.1086/491729}, \href
  {http://adsabs.harvard.edu/abs/2005ApJ...634..442S} {634, 442}

\bibitem[\protect\citeauthoryear{{Schnee}, {Bethell}  \& {Goodman}}{{Schnee}
  et~al.}{2006}]{2006ApJ...640L..47S}
{Schnee} S.,  {Bethell} T.,   {Goodman} A.,  2006, \mn@doi [\apjl]
  {10.1086/503292}, \href {http://adsabs.harvard.edu/abs/2006ApJ...640L..47S}
  {640, L47}

\bibitem[\protect\citeauthoryear{{Schnee}, {Kauffmann}, {Goodman}  \&
  {Bertoldi}}{{Schnee} et~al.}{2007}]{2007ApJ...657..838S}
{Schnee} S.,  {Kauffmann} J.,  {Goodman} A.,   {Bertoldi} F.,  2007, \mn@doi
  [\apj] {10.1086/511054}, \href
  {http://adsabs.harvard.edu/abs/2007ApJ...657..838S} {657, 838}

\bibitem[\protect\citeauthoryear{{Schnee}, {Li}, {Goodman}  \&
  {Sargent}}{{Schnee} et~al.}{2008}]{2008ApJ...684.1228S}
{Schnee} S.,  {Li} J.,  {Goodman} A.~A.,   {Sargent} A.~I.,  2008, \mn@doi
  [\apj] {10.1086/590375}, \href
  {http://adsabs.harvard.edu/abs/2008ApJ...684.1228S} {684, 1228}

\bibitem[\protect\citeauthoryear{{Schruba} et~al.,}{{Schruba}
  et~al.}{2011}]{SCHRUBA11}
{Schruba} A.,  et~al., 2011, \mn@doi [\aj] {10.1088/0004-6256/142/2/37}, \href
  {http://adsabs.harvard.edu/abs/2011AJ....142...37S} {142, 37}

\bibitem[\protect\citeauthoryear{{Schruba} et~al.,}{{Schruba}
  et~al.}{2012}]{SCHRUBA12}
{Schruba} A.,  et~al., 2012, \mn@doi [\aj] {10.1088/0004-6256/143/6/138}, \href
  {http://adsabs.harvard.edu/abs/2012AJ....143..138S} {143, 138}

\bibitem[\protect\citeauthoryear{{Shetty}, {Glover}, {Dullemond}  \&
  {Klessen}}{{Shetty} et~al.}{2011}]{SHETTY11}
{Shetty} R.,  {Glover} S.~C.,  {Dullemond} C.~P.,   {Klessen} R.~S.,  2011,
  \mn@doi [\mnras] {10.1111/j.1365-2966.2010.18005.x}, \href
  {http://adsabs.harvard.edu/abs/2011MNRAS.412.1686S} {412, 1686}

\bibitem[\protect\citeauthoryear{{Skibba} et~al.,}{{Skibba}
  et~al.}{2012}]{SKIBBA12}
{Skibba} R.~A.,  et~al., 2012, \mn@doi [\apj] {10.1088/0004-637X/761/1/42},
  \href {http://adsabs.harvard.edu/abs/2012ApJ...761...42S} {761, 42}

\bibitem[\protect\citeauthoryear{{Stanimirovic}, {Staveley-Smith}, {van der
  Hulst}, {Bontekoe}, {Kester}  \& {Jones}}{{Stanimirovic}
  et~al.}{2000}]{STANIMIROVIC00}
{Stanimirovic} S.,  {Staveley-Smith} L.,  {van der Hulst} J.~M.,  {Bontekoe}
  T.~R.,  {Kester} D.~J.~M.,   {Jones} P.~A.,  2000, \mn@doi [\mnras]
  {10.1046/j.1365-8711.2000.03480.x}, \href
  {http://adsabs.harvard.edu/abs/2000MNRAS.315..791S} {315, 791}

\bibitem[\protect\citeauthoryear{{Sternberg}, {Le Petit}, {Roueff}  \& {Le
  Bourlot}}{{Sternberg} et~al.}{2014}]{STERNBERG14}
{Sternberg} A.,  {Le Petit} F.,  {Roueff} E.,   {Le Bourlot} J.,  2014, \mn@doi
  [\apj] {10.1088/0004-637X/790/1/10}, \href
  {http://adsabs.harvard.edu/abs/2014ApJ...790...10S} {790, 10}

\bibitem[\protect\citeauthoryear{{Welty}, {Xue}  \& {Wong}}{{Welty}
  et~al.}{2012}]{WELTY12}
{Welty} D.~E.,  {Xue} R.,   {Wong} T.,  2012, \mn@doi [\apj]
  {10.1088/0004-637X/745/2/173}, \href
  {http://adsabs.harvard.edu/abs/2012ApJ...745..173W} {745, 173}

\bibitem[\protect\citeauthoryear{{Westerlund}}{{Westerlund}}{1997}]{WESTERLUND97}
{Westerlund} B.~E.,  1997, {The Magellanic Clouds}

\bibitem[\protect\citeauthoryear{{Wilson}}{{Wilson}}{1995}]{WILSON95}
{Wilson} C.~D.,  1995, \mn@doi [\apjl] {10.1086/309615}, \href
  {http://adsabs.harvard.edu/abs/1995ApJ...448L..97W} {448, L97}

\bibitem[\protect\citeauthoryear{{Wolfire}, {Hollenbach}  \& {McKee}}{{Wolfire}
  et~al.}{2010}]{WOLFIRE10}
{Wolfire} M.~G.,  {Hollenbach} D.,   {McKee} C.~F.,  2010, \mn@doi [\apj]
  {10.1088/0004-637X/716/2/1191}, \href
  {http://adsabs.harvard.edu/abs/2010ApJ...716.1191W} {716, 1191}

\bibitem[\protect\citeauthoryear{{Wong} et~al.,}{{Wong} et~al.}{2011}]{WONG11}
{Wong} T.,  et~al., 2011, \mn@doi [\apjs] {10.1088/0067-0049/197/2/16}, \href
  {http://adsabs.harvard.edu/abs/2011ApJS..197...16W} {197, 16}

\bibitem[\protect\citeauthoryear{{Zaritsky}, {Harris}, {Thompson}  \&
  {Grebel}}{{Zaritsky} et~al.}{2004}]{ZARITSKY04}
{Zaritsky} D.,  {Harris} J.,  {Thompson} I.~B.,   {Grebel} E.~K.,  2004,
  \mn@doi [\aj] {10.1086/423910}, \href
  {http://adsabs.harvard.edu/abs/2004AJ....128.1606Z} {128, 1606}

\bibitem[\protect\citeauthoryear{{van Dishoeck} \& {Black}}{{van Dishoeck} \&
  {Black}}{1988}]{VB88}
{van Dishoeck} E.~F.,  {Black} J.~H.,  1988, \mn@doi [\apj] {10.1086/166877},
  \href {http://adsabs.harvard.edu/abs/1988ApJ...334..771V} {334, 771}

\makeatother
\end{thebibliography}
\end{document}